\documentclass[3p,final,11pt,authoryear]{elsarticle}
\usepackage[utf8]{inputenc}
\usepackage{graphicx}
\usepackage{mathtools}
\usepackage{amsmath}
\usepackage{lscape}
\usepackage[colorlinks]{hyperref}
\usepackage[nameinlink,noabbrev]{cleveref}
\usepackage{geometry}
\usepackage{enumitem}
\usepackage{natbib}
\usepackage{subcaption}
\usepackage[tableposition = top]{caption}
\usepackage{booktabs}
\usepackage{algorithm}
\usepackage{algorithmic}
\usepackage{hyperref}
\usepackage{float}
\usepackage{graphicx,subcaption}
\usepackage{caption}
\usepackage{xcolor}
\usepackage{lineno}
\usepackage{multirow}
\usepackage{soul}
\usepackage{tabularx}
\usepackage{adjustbox}
\usepackage{rotating}
\usepackage{lscape}

\setcitestyle{authoryear,open={(},close={)}}
\usepackage{setspace}

\begin{document}

\begin{frontmatter}

\title{Decoding Pedestrian Stress on Urban Streets using Electrodermal Activity Monitoring in Virtual Immersive Reality}

    \author[1]{Mohsen Nazemi}
    \author[1]{Bara Rababah}
    \author[1]{Daniel Ramos}
    \author[1]{Tangxu Zhao}
    \author[1]{Bilal Farooq\corref{cor1}}\ead{bilal.farooq@torontomu.ca}
    
    \address[1]{Laboratory of Innovations in Transportation (LiTrans), Toronto Metropolitan University, Canada}
    \cortext[cor1]{Corresponding Author.}

\begin{abstract}

The pedestrian stress level is shown to significantly influence human cognitive processes and, subsequently, decision-making, e.g., the decision to select a gap and cross a street. This paper systematically studies the stress experienced by a pedestrian when crossing a street under different experimental manipulations by monitoring the Electrodermal Activity (EDA) using the Galvanic Skin Response (GSR) sensor. To fulfil the research objectives, a dynamic and immersive virtual reality (VR) platform was used, which is suitable for eliciting and capturing pedestrian's emotional responses in conjunction with monitoring their EDA. A total of 171 individuals participated in the experiment, tasked to cross a two-way street at mid-block with no signal control. Mixed effects models were employed to compare the influence of socio-demographics, social influence, vehicle technology, environment, road design, and traffic variables on the stress levels of the participants. The results indicated that having a street median in the middle of the road operates as a refuge and significantly reduced stress. Younger participants (18-24 years) were calmer than the relatively older participants (55-65 years). Arousal levels were higher when it came to the characteristics of the avatar (virtual pedestrian) in the simulation, especially for those avatars with adventurous traits. The pedestrian location influenced stress since the stress was higher on the street while crossing than waiting on the sidewalk. Significant causes of arousal were fear of accidents and an actual accident for pedestrians. The estimated random effects show a high degree of physical and mental learning by the participants while going through the scenarios. 

\end{abstract}

\end{frontmatter}


\section{Introduction}
\label{sec:Introduction}

{Pedestrian comfort and safety are regularly cited as the top priorities for residents, elected officials, and staff in cities all over the world, while the Vision Zero approach emphasizes no loss of life as a result of interactions with traffic \citep{cityoftorontoFatalitiesVisionZero2023, varnild2020types, kronenberg2015achieving}. The majority of pedestrian fatalities occur on urban roads and pedestrian collisions more often occur near intersections when pedestrians are crossing a roadway \citep{transportcanadaCanadianMotorVehicle2023, torontopoliceservicePublicSafetyData2022}. Additionally, the large block sizes and absence of pedestrian crossovers encourage pedestrians to jaywalk across streets, resulting in mid-block areas becoming the second most common locations for pedestrian fatalities, accounting for 14\% \citep{statisticscanadaCircumstancesSurroundingPedestrian2023}. In 2021, 84\% of pedestrian fatalities in the United States were in urban locations, while 75\% of these incidents occurred at non-intersections \citep{national2023pedestrians}. While U.S. pedestrian fatalities increased by 19\% per capita from 2010 to 2018, these fatality rates remain much higher than those in the UK, where per-capita fatalities rose by 4\%. In Germany, 74\% of pedestrian fatalities occurred on urban roads in 2019 \citep{european2021facts}, whereas in the UK, between 2018 and 2022, 66\% of pedestrian fatalities occurred on urban roads \citep{department2023reported}.}

As Automated Vechiles (AVs) become more prevalent, a key concern arises as pedestrians no longer have the explicit communication techniques they used to have with drivers when negotiating the road with AVs. Such discrepancies might exacerbate current pedestrian stress and therefore, urge for studies to explore pedestrian behaviour in more detail. Various environmental factors, such as time of day and weather conditions, along with street design, have also been shown to impact pedestrian decision-making behaviour. Furthermore, pedestrians may influence each other's behaviour under the social conformity theory.

Given the complexities of pedestrian decision-making when negotiating the road with other road users, there is a strong need to enhance our understanding of pedestrian behaviour when crossing a street. Emotional state, especially stress level, is shown to influence human cognitive processes and subsequently decision-making, e.g., the decision to select a gap \citep{paschalidisModellingEffectsStress2018} and cross a street \citep{farooq2022}. Therefore, it is crucial to understand how stress levels influence pedestrian decisions while interacting with traffic in a range of different conditions. Particularly, there is a need to investigate the effects of AVs on pedestrian stress levels and decision-making processes under varying road-crossing scenarios, traffic mix, environment, geometry, social influence, and socio-demographics. Physiological sensors have the capacity to uncover new knowledge and provide a deeper understanding of human emotions. These sensors work best in controlled laboratory environments, especially when combined with Virtual Reality (VR). At the same time, VR provides a suitable platform to reproduce scenarios and safety study human-machine interactions \citep{tranReviewVirtualReality2021}. Various recent studies have established the efficacy and validity of VR in studying pedestrian behaviour, especially in terms of ecological, face, content, and construct validity \citep{farooq2024workshop, deb2017efficacy}. 

This research studied the experienced stress of pedestrians when crossing an unsignalized two-lane urban street. To fulfil the research objectives, an interactive and immersive VR platform was used, which is suitable for eliciting and capturing pedestrian emotional responses and monitoring their EDA, using a GSR sensor. The research design incorporated variations in vehicle type (i.e., normal vehicles, AVs with roof sign, AVs with  external human–machine interface (eHMI)), avatar's behaviour (i.e., no avatar, standing avatar, conservative avatar, and adventurous avatar), traffic flow (i.e., low arrival rate and high speed, and high arrival rate and high speed), street design (i.e., median), time of day (i.e., day and night), and weather conditions (i.e., sunny, rainy, and snowy). This comprehensive approach ensures the analyses of all critical factors contributing to pedestrian stress in real-world conditions. The research questions addressed in this research are (1) Do different vehicle types and signals portrayed to pedestrians impact their incurred crossing stress? (2) Does the presence of other pedestrians and their behaviour influence the incurred stress of a pedestrian? and (3) Do environmental, traffic, and geometric variables such as traffic flow, median, time of day, and weather conditions impact the stress level of a pedestrian? Mixed effects models were employed to compare the influence of independent variables on the stress levels of the participants. The key contributions of this research include:

\begin{itemize}
    \item Design and development of a full-scale immersive VR walking experiment with a total of 171 participants to study pedestrian stress in a realistic and controlled environment;
    \item Analyzing the resulting multi-modal data to quantify the experienced stress under different experimental manipulations that have  rarely been studied before;
    \item Detailed contextualized assessment of pedestrian emotional responses to identify the sources of stress and their influence.
\end{itemize}

The rest of the paper is organized as follows. Section \ref{sec:Background} summarizes the previous research in relation to the interaction of pedestrians and selected variables of the present study and the applied research methods. Section \ref{sec:Methods} describes the experiment setup, the experimental design, and the participants of the experiment. Section \ref{methods} presents the modelling approaches to quantify the relationship between stress and various variables. Section \ref{sec:Results} presents the results of the measurement variable in relation to the desired experiment factors. Finally, section \ref{sec:conclusions} concludes the paper by summarizing this study’s innovations, contributions, and limitations, and by providing direction for future studies.

\section{Background}
\label{sec:Background}
To the best of our knowledge, only a handful of studies exist that analyzed the stress experienced by a pedestrian on urban roads, therefore, here we provide a more general background on the state of research related to pedestrian behaviour and the associated topics.

\subsection{Virtual Reality and Pedestrian Behaviour}
In recent years, VR has become an important tool for studying pedestrian behaviour, especially in terms of crossing \citep{kwon2022pedestrians}, way-finding \citep{van2024comparison}, evacuation \citep{ronchi2015evacuation}, distraction \citep{sobhaniDistractedPedestriansCrossing2017}, and interaction with AVs \citep{kalatianDecodingPedestrianAutomated2021}. VR provides a safe and controlled environment to study the influence of a range of variables and scenarios, though the high entry barrier in terms of access to the required hardware and software development skills remains. \cite{farooq2018virtual} concluded that in comparison to text and video-based tools, VR can better capture the heterogeneity in pedestrian behaviour. \cite{deb2017efficacy} established the efficacy of using VR in pedestrian research by comparing the average walking speed in VR to that of published real-world norms and found no difference. Various recent studies have established the validity of VR in studying pedestrian behaviour, especially in terms of ecological, face, content, and construct validity \citep{farooq2024workshop}.

A major body of literature studied the crossing behaviour of pedestrian through measuring kinematic variables such as wait time \citep{kalatianDecodingPedestrianAutomated2021}, critical gap acceptance \citep{zhaoInvisibleGorillaPedestrianAV2023}, reaction time \citep{zhuNovelAgentbasedFramework2021}, time to collision \citep{rasouliAgreeingCrossHow2017, stadlerToolNotToy2019a}, and Post Encroachment Time \citep{sobhaniDistractedPedestriansCrossing2017}. Frequently, variables such as road geometry and features, weather conditions, gender, and age were considered as additional predictors to construct crossing behaviour models. Most of these studies have been conducted at unsignalized pedestrian crosswalks with the objective to model the pedestrian decision-making processes for crossing under different conditions. Furthermore, it is noted that these studies, especially the ones that used VR, generally rely on data from only a few tens of participants. 

\subsection{Impact of Social Influence}

The social influence in terms of the effect of nearby pedestrians' behaviour on a pedestrian has been studied in previous research. Imitation and leader/follower effects (i.e. some pedestrians may follow the crossing choices of others, while others may prompt their company to a specific behaviour) is a commonly observed behaviour in groups of pedestrians waiting to cross a street \citep{papadimitriouIntroducingHumanFactors2016, tezcanPedestrianCrossingBehavior2019}. Pedestrians contemplating to cross in a group use \textit{social information} to judge a safe gap in traffic. Two examples of social information in the context of crossing a street are as follows. In a group of pedestrians waiting to cross a street, the immediate adjacent pedestrians to the one who crosses first (the leader), tend to cross before other waiting pedestrians and they are more ready to follow the behaviour of the leader in the same group. There are pedestrians that start to cross, but then return back to the sidewalk. These individuals also use social information, however, they make incorrect decisions about the timing of their crossing \citep{fariaCollectiveBehaviorRoad2010}. Even though studies have identified the reasoning behind such pedestrian behaviour, there still exist many unjustified observations. For example, it has been shown that male pedestrians tend to leave the platoon of pedestrians waiting to cross a street, but they are not willing to perform risky crossing behaviour when they are alone \citep{tezcanPedestrianCrossingBehavior2019}. Or, male pedestrians tend to follow others more than female pedestrians in a platoon of waiting pedestrians \citep{fariaCollectiveBehaviorRoad2010}.
It was observed that on average, a pedestrian is 1.5 to 2.5 times more likely to cross if his/her neighbour starts to cross \citep{fariaCollectiveBehaviorRoad2010}. Another research found that participants who had an adventurous partner crossed with a smaller gap on average than those who had a conservative partner \citep{y.jiangJointActionVirtual2019}. It was found that waiting pedestrians encourage others to wait, too, and adventurous pedestrians who cross at a red light encourage others to cross, as well \citep{rakotoariveloIntroducingSocialInfluence2020}. Gender and age were found to have different influences on other pedestrians. For example, compared to adults, young teenagers accepted riskier gaps when crossing with a friend than when doing so alone \citep{onealHowDoesCrossing2019}. The presence of others influences the decisions of women more than men \citep{yagilBeliefsMotivesSituational2000}, and women were more likely to follow others than men \citep{fariaCollectiveBehaviorRoad2010}. It has been also found that social influence may lead to hesitations and incorrect decisions \citep{coeugnetRisktakingEmotionsSociocognitive2019}. {\cite{kinateder2014social} used VR to study the effects of social influence in evacuation behaviour during tunnel fires. The study concluded that social influence modulated the likelihood of adequate safety behaviour during the evacuation.}  Overall, studies agree that other pedestrians influence other pedestrians' behaviour, but disagree on whether the presence of others encourages riskier or less risky behaviour \citep{predhumeauPedestrianBehaviorShared2023}. 

Many studies have looked into simulated pedestrian (also referred to as avatar or agent) activity in the simulated environment. \cite{kwon2015} used a randomized block design to expose each block of participants to virtual avatars with varying levels of realism: realistic 3D human avatars, cartoon-like 3D avatars, and human pictures. The levels of social anxiety were assessed using eye avoidance behaviour, which is a common sign of anxiety. They found that virtual avatars might affect how the participants rated the avatar's impact. When an avatar was not present, pedestrians perceived the degree of influence by other people differently, suggesting that engaging avatars did make a difference in influencing arousal levels compared to the avatars doing nothing. Additionally, \cite{pan2015} looked at how people imitate virtual avatars. They discovered that when participants were required to follow avatars, they made far fewer mistakes than when they were required to follow a ball. The movements made by the participants' hands were imitations of the hands of the virtual character, indicating that there was a high degree of similarity in how the human and virtual avatar hands behaved compared to the virtual balls and human hands. On the other hand, \cite{latoschik2017} tested how user-friendly or approachable avatars are in comparison to real people and found that avatars were less user-friendly and less approachable than humans. The difference shows how the interaction with avatars could be complex and that further research regarding the specific features of avatars and their design in such virtual environments is required when assessing their impact on pedestrian arousal levels.

\subsection{Pedestrian-AV Interactions}

The introduction of AVs has the potential to bring about many challenges and uncertainties, especially for pedestrians who are among the most vulnerable road users. As AVs become more common, a potential challenge lies in the negotiation of right-of-way between AVs and pedestrians in conflict zones, which could potentially increase injuries, fatalities, or at least pedestrian hesitation \citep{sottileUberSelfdrivingCar2023}. 
Previous research has shown that pedestrians have diverse and imperfect crossing behaviour in the presence of AVs. Their design must consider this variety of behaviours and follow socially compliant rules in order to be understood and accepted by pedestrians \citep{predhumeauPedestrianBehaviorShared2023}. The communication between a vehicle and a pedestrian can happen in an explicit or implicit manner. Explicit communication includes direct messages exchanged between road users, e.g., signal and sound, while implicit communication implies indirect messages in which the content is not directly addressed, e.g., reducing speed to encourage pedestrians to cross. \citep{fuestTaxonomyTrafficSituations2018}.

Traditionally, pedestrians seek explicit communication techniques such as eye-contact with drivers, honking, flashing headlights by drivers, or hand gestures, as well as implicit techniques, such as vehicle motion cues, when contemplating a crossing \citep{leeRoadUsersRarely2021, rasouliAgreeingCrossHow2017}. To compensate for the absence of eye-contact and increase safety, AV manufacturers consider developing visual external human–machine interfaces for AVs \citep{stadlerToolNotToy2019a}. However, with the rising number of eHMI design concepts emerging in both academic and industrial settings, the ability of eHMIs to handle complex scenarios has become a pivotal factor influencing its widespread adoption. While the effectiveness of visual eHMI designs has been debated extensively \citep{limHowDesignEHMI2023, limUIDesignEHMI2022, bazilinskyySurveyEHMIConcepts2019, clamannEvaluationVehicletopedestrianCommunication2017, o.benderiusBestRatedHuman2018, stadlerToolNotToy2019a}, there is a line of research that found pedestrian crossing behaviour is highly influenced by the implicit cues, rather than the explicit ones \citep{leeRoadUsersRarely2021, mooreCaseImplicitExternal2019}. 

The review of the literature shows that the interactions between pedestrians and vehicular traffic have predominantly been treated as a rather mechanistic process. There is a need to better understand the variability of pedestrian behaviour using advanced methods. In this regard, recent studies suggest frameworks based on cognitive and affective approaches to human behaviour \citep{predhumeauPedestrianBehaviorShared2023}. For this purpose, physiological sensors---and in particular GSR sensors---offer quantitative measurements of an individual's affective states when performing different tasks, which might be able to reveal causal relationships behind human behaviour \citep{Boucsein_2012}.

\subsection{Pedestrian Stress}
The experienced stress level has a strong influence on human behaviour and decision-making process \citep{farooq2022, paschalidisModellingEffectsStress2018}. The existing literature has primarily examined the self-reported perceived stress of pedestrians in various situations, e.g., when interacting with other road users \citep{specktor2023perceptions, zheng2017joint} or evaluating the future street scenarios \citep{argota2024virtual}. While the self-reported stress level on a Likert scale has been regularly used as a proxy for experienced stress, such indicators may suffer from issues such as response bias, social conformity bias, lack of consideration for cultural differences, bias due to range and direction of scale, and lack of context \citep{tourangeau2000psychology}.

On the other hand, only a handful of studies exist that measured the stress experienced by pedestrians using neurophysiological sensors. \cite{kobayashi2020examination} examined the effect of different geometric designs of blindspots in an intersection on pedestrian stress. For this purpose, the study used a GSR sensor to measure the stress of 30 participants in five different design configurations. \cite{lajeunesse2021measuring} developed a naturalistic walking study where 15 pedestrians wore an instrumented wristband and GPS recorder for their walking trips for one week. The biosensors in the wristband recorded the EDA and blood volume pulse, which were used to quantify the experienced stress. The study reported that participants experienced higher stress around busy streets and in areas with industrial and mixed land uses. They experienced lower stress in low-density residential areas, forests, parks, and university campus environments. \cite{MudassarKalatianFarooq2022} used the GSR sensor in VR to measure the temporal patterns of stress for each participant. The study analyzed the relationship between stress and pedestrian's kinematic indicators, including speed, acceleration and distance from the vehicle and found strong correlations. \cite{farooq2022} developed an ordered-logit model for experienced average stress levels to quantify the effects of various variables in the context of mid-block crossing. For this purpose, the stress measurements collected via a GSR in VR were discretized into ordered categories (low, medium, and high). While such an approach provides a useful start in terms of understanding pedestrian stress, unfortunately, the lack of granularity misses out on various important spatio-temporal variations, e.g., a sudden rise in instantaneous stress level due to an approaching vehicle mid-crossing.

\subsection{{Current Research Gaps}}
{In the recent literature, there are only a few studies that measured the stress experienced by pedestrian when interacting with other road users. Furthermore, the focus of these studies has been limited to either understanding the individual stress profile or aggregate effects of a limited number of variables only. A systematic and empirical analysis of the contextual changes in experienced stress due to pedestrian location (e.g., being on a sidewalk or in the middle of the road), common events (e.g., approaching vehicle), and rare events (e.g., near-miss collision experience) are also missing from the existing literature. Previous studies on pedestrian behaviour in normal as well as panic situations have shown that the social influence play a significant role in influencing the pedestrian decision making. However, to the best of our knowledge, the direct relationship between social influence and experienced stress has not been empirically studied. Finally, most of the existing studies that used VR to study the pedestrian behaviour collected data on only a limited number of homogeneous respondents. For the generalizability of the results and more in-depth and integrated analysis, there is a strong need for conducting large-scale VR data gathering campaigns that go beyond the minimum requirements of power analysis.}

\section{Experiment}
\label{sec:Methods}
To study experienced stress and walking behaviour, we developed a detailed controlled experiment that continuously monitored the participant's EDA using a GSR sensor and motion in an immersive and dynamic VR environment. In this section, we describe several important components of the experiment. 

\subsection{Considered Factors}

After reviewing the previous studies, key factors influencing pedestrian crossing behaviour were identified. Particularly, the research incorporated five distinct types of factors aimed at manipulating diverse environments for the participants. Table \ref{tab:tabexperimentfactors} shows these factors, the associated variables, and variable levels. As vehicle type, i.e., normal car, AV, and AV communication interface were found to have an influence on pedestrian behaviour \citep{leeRoadUsersRarely2021, stadlerToolNotToy2019a}, three vehicle types and vehicle appearances were designed, including AVs with eHMIs. The inconclusive results of the previous research on repeating the adventurous behaviour of other pedestrians \citep{predhumeauPedestrianBehaviorShared2023} encouraged including this phenomenon in this study. 
In this context, three behaviours of the other pedestrians were investigated, i.e., 1) a standing pedestrian simply waiting on the sidewalk, {2) a conservative pedestrian crossing the street in a cautious manner at a regular walking speed of 1 m/sec (for an example, see the video: \url{https://www.youtube.com/watch?v=s8SYBGwT968}), and 3) an adventurous pedestrian who chose a more daring approach of running to cross the street with a speed of 2 m/sec (for an example, see the video: \url{https://www.youtube.com/watch?v=ZB2OIgw2L1k})}. 
The remaining factors were influenced by street medians and various environmental factors. Since it was objectively shown that pedestrian deaths and injuries increase with traffic volumes and decrease with the presence of raised medians, these two variables were also included in this study to explore the arousal level of the participants under these interventions \citep{dumbaughMostVulnerableUser2024}. Furthermore, recent studies showed time of day and seasonal patterns on the injury rate and injury severity of pedestrians \citep{akhtarDiurnalSeasonalRelationships2024}. Therefore, these variables were considered in the study design with appropriate levels to create meaningful variations between different scenarios. Additionally, we also collected the age and gender information of the participants using an associated online survey.

{The traffic generator model used in this research takes three inputs to simulate a uniform traffic flow: 1) arrival rate [vehicles/hour], 2) maximum vehicle speed [km/hour], and 3) vehicle spacing [m]. Various input values were tested to ensure that the generated conditions were reasonable, replicating realistic traffic flow patterns. Ultimately, the input variables were adjusted to create two scenarios: 1) high arrival rate and low speed, and 2) low arrival rate and high speed.}

{The traffic flow characteristics for each of these scenarios were measured and are summarized in Table \ref{tab:tabexperimentfactors}. It is important to note that these measurements apply to scenarios where no conservative or adventurous avatars are present, as their inclusion would create shockwaves, altering the traffic flow characteristics. Furthermore, the traffic conditions were consistently replicated across participants, demonstrating the reproducibility feature of the VR experiments.}

\begin{table}[htbp]
  \centering
  \caption{Experiment factors, variables, and variable levels}
  \resizebox{\textwidth}{!}{
    \begin{tabular}{llll}
    \toprule
    \multicolumn{1}{p{11.25em}}{Factor} & \multicolumn{1}{p{11.25em}}{Variable} & \multicolumn{2}{p{32.5em}}{Levels} \\
    \midrule
    \midrule
    \multicolumn{1}{p{11.25em}}{Vehicle type} & \multicolumn{1}{p{11.25em}}{Vehicle appearance} & \multicolumn{2}{p{32.5em}}{Normal car, AV with roof sign, AV with signal boards} \\
    \midrule
    \multicolumn{1}{p{11.25em}}{Social influence} & \multicolumn{1}{p{11.25em}}{Avatar behaviour} & \multicolumn{2}{l}{No avatar, Standing avatar, Conservative avatar, Adventurous avatar} \\
    \midrule
          &       & \multicolumn{1}{p{16.25em}}{High arrival rate and low speed} & \multicolumn{1}{p{16.25em}}{Low arrival rate and high speed} \\
\cmidrule{3-4}          & \multicolumn{1}{p{11.25em}}{Traffic flow [veh/hr]} & 1200  & 1113 \\
    \multicolumn{1}{p{11.25em}}{Traffic} & \multicolumn{1}{p{11.25em}}{Average speed [km/hr]} & 20    & 40 \\
    \multicolumn{1}{p{11.25em}}{characteristics} & \multicolumn{1}{p{11.25em}}{Average gap [s]} & 3.0   & 3.2 \\
          & \multicolumn{1}{p{11.25em}}{Average clearance [m]} & 16.7  & 35.9 \\
          & \multicolumn{1}{p{11.25em}}{Vehicle share} & 0\% AV, 100\% AV & 0\% AV, 100\% AV \\
    \midrule
    \multicolumn{1}{p{11.25em}}{Road type} & \multicolumn{1}{p{11.25em}}{Street design} & \multicolumn{2}{l}{No median, With median} \\
    \midrule
    \multicolumn{1}{p{11.25em}}{Environmental} & \multicolumn{1}{p{11.25em}}{Time of day} & \multicolumn{2}{l}{Day, Night} \\
    \multicolumn{1}{p{11.25em}}{characteristics} & \multicolumn{1}{p{11.25em}}{Weather} & \multicolumn{2}{l}{Clear, Rain, Snow} \\
    \midrule
          &       &       &  \\
    \multicolumn{4}{l}{Notes:} \\
    \multicolumn{4}{l}{Gap is a measure of the time between the rear bumper of the first vehicle and the front bumper of the second vehicle.} \\
    \multicolumn{4}{l}{Clearance is the distance between the rear bumper of the leading vehicle and the front bumper of the following vehicle.} \\
    \end{tabular}%
    }
  \label{tab:tabexperimentfactors}%
\end{table}%


\subsection{Virtual Reality Environment}
\begin{figure}[!ht] 
\centering

    \begin{subfigure}{0.47\textwidth}
        \includegraphics[width=\textwidth]{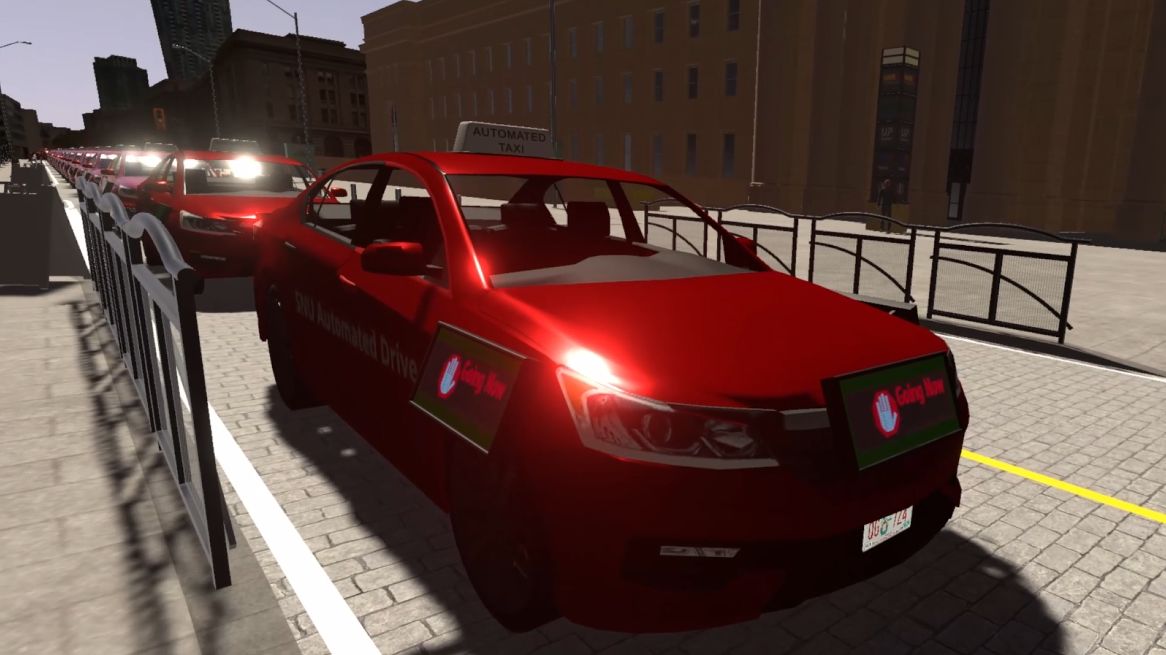}
        \caption{AV with eHMI signalling to stop (in red)}
        \label{fig:av}
    \end{subfigure}%
    \hspace{0.02\textwidth}
    \begin{subfigure}{0.47\textwidth}
        \includegraphics[width=\textwidth]{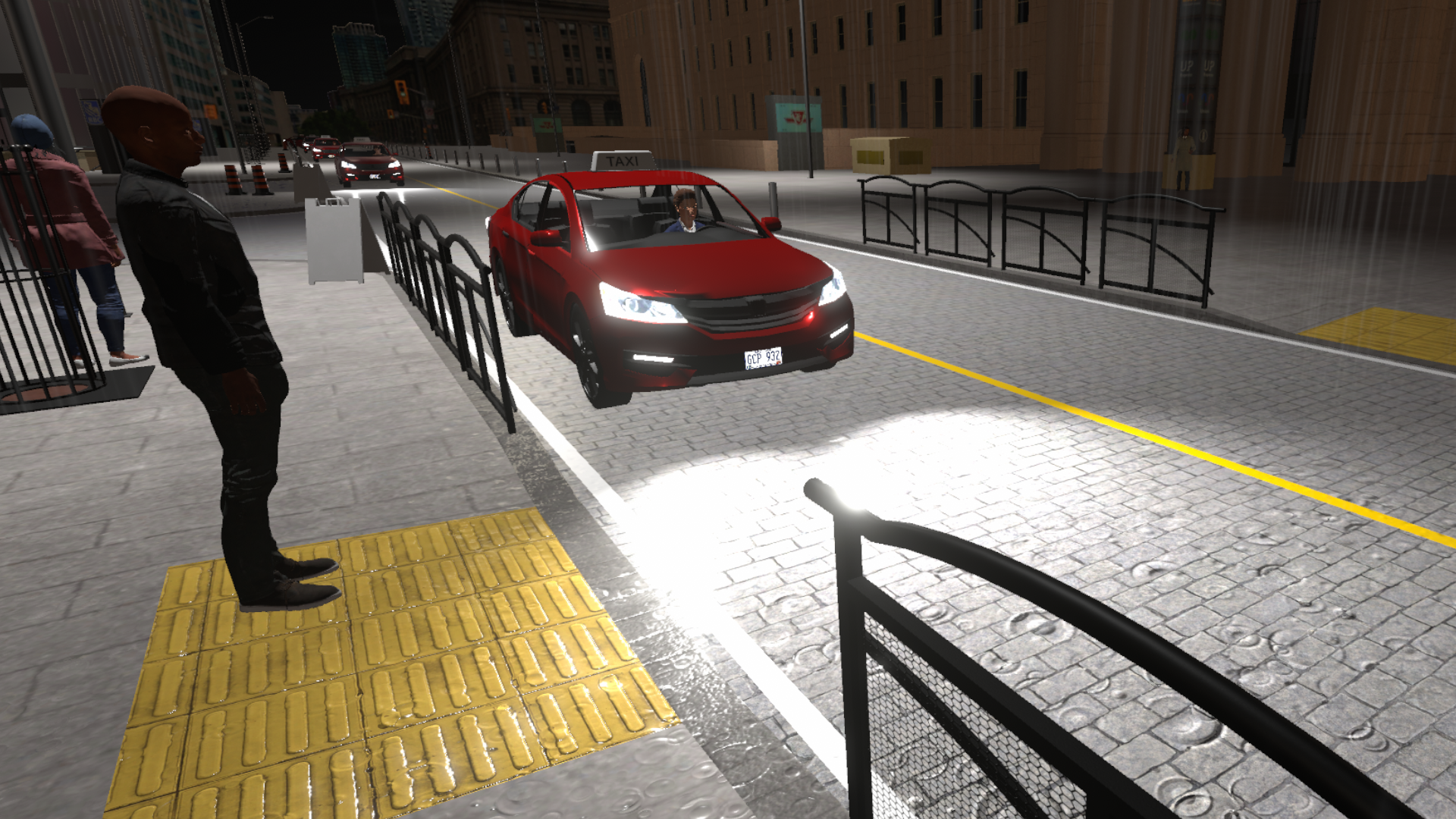} 
        \caption{Avatar responding to the AV eHMI signal and stopping}
        \label{fig:agent}
    \end{subfigure}
    \begin{subfigure}{0.47\textwidth}
        \includegraphics[width=\textwidth]{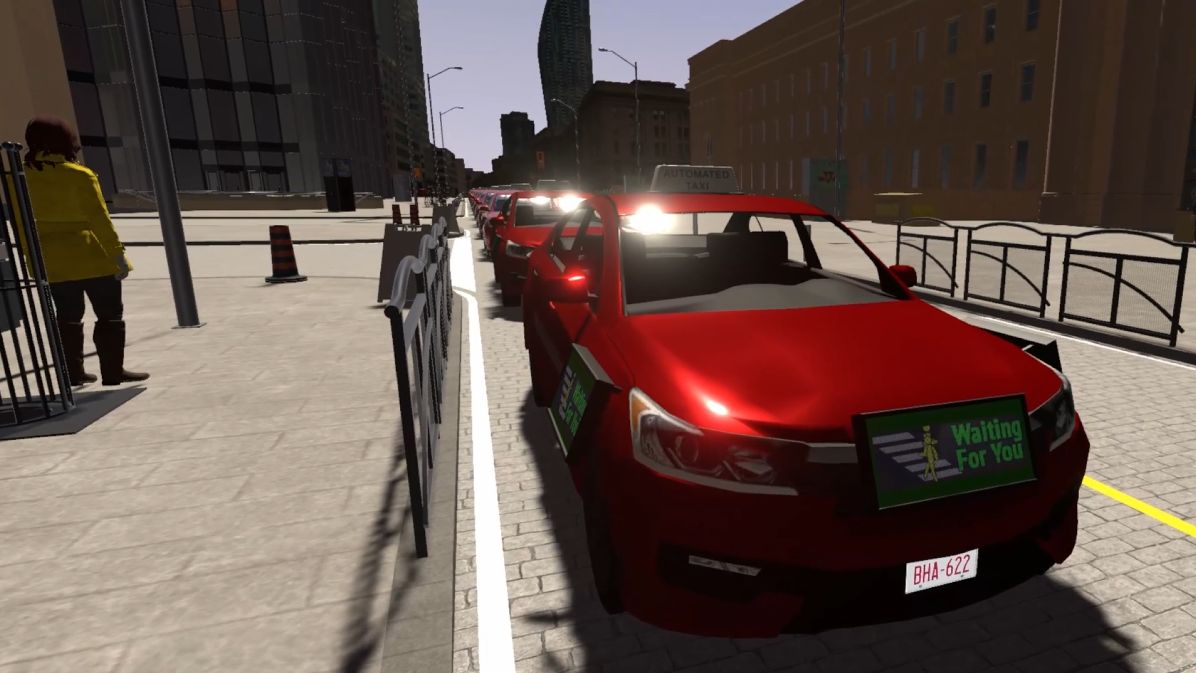}
        \caption{AV with eHMI signalling to go (in green)}
        \label{fig:c}
    \end{subfigure}%
    \hspace{0.02\textwidth}
    \begin{subfigure}{0.47\textwidth}
        \includegraphics[width=\textwidth]{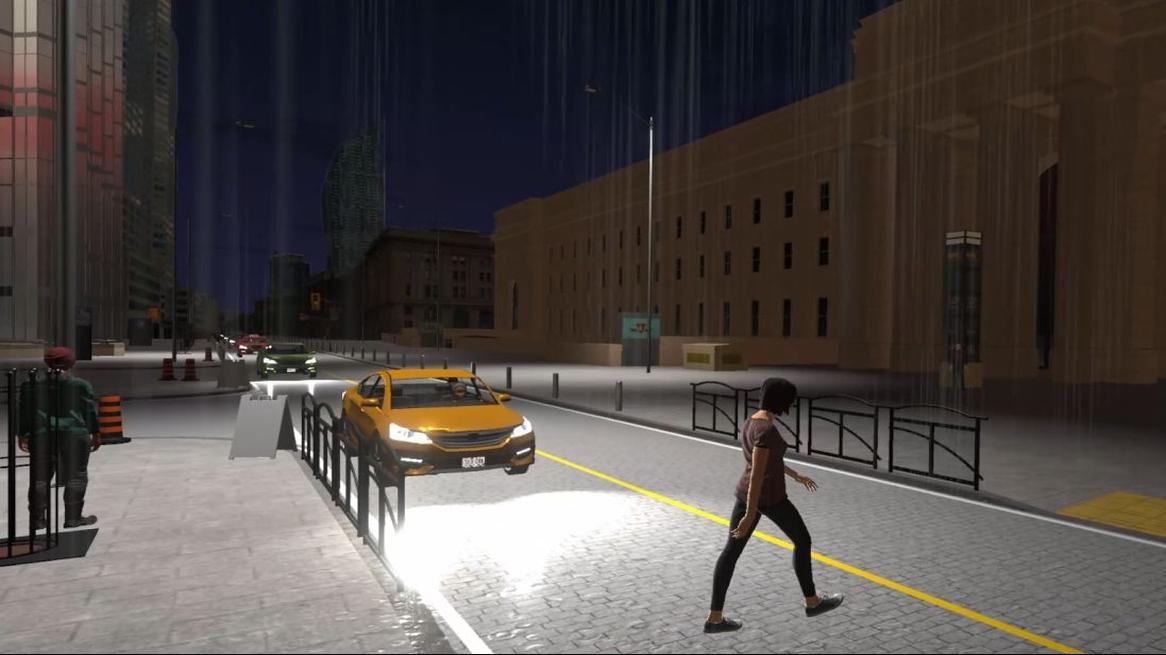}
        \caption{Avatar responding to the AV eHMI signal and walking}
        \label{fig:avago}
    \end{subfigure}
    \begin{subfigure}{0.47\textwidth}
        \includegraphics[width=\textwidth]{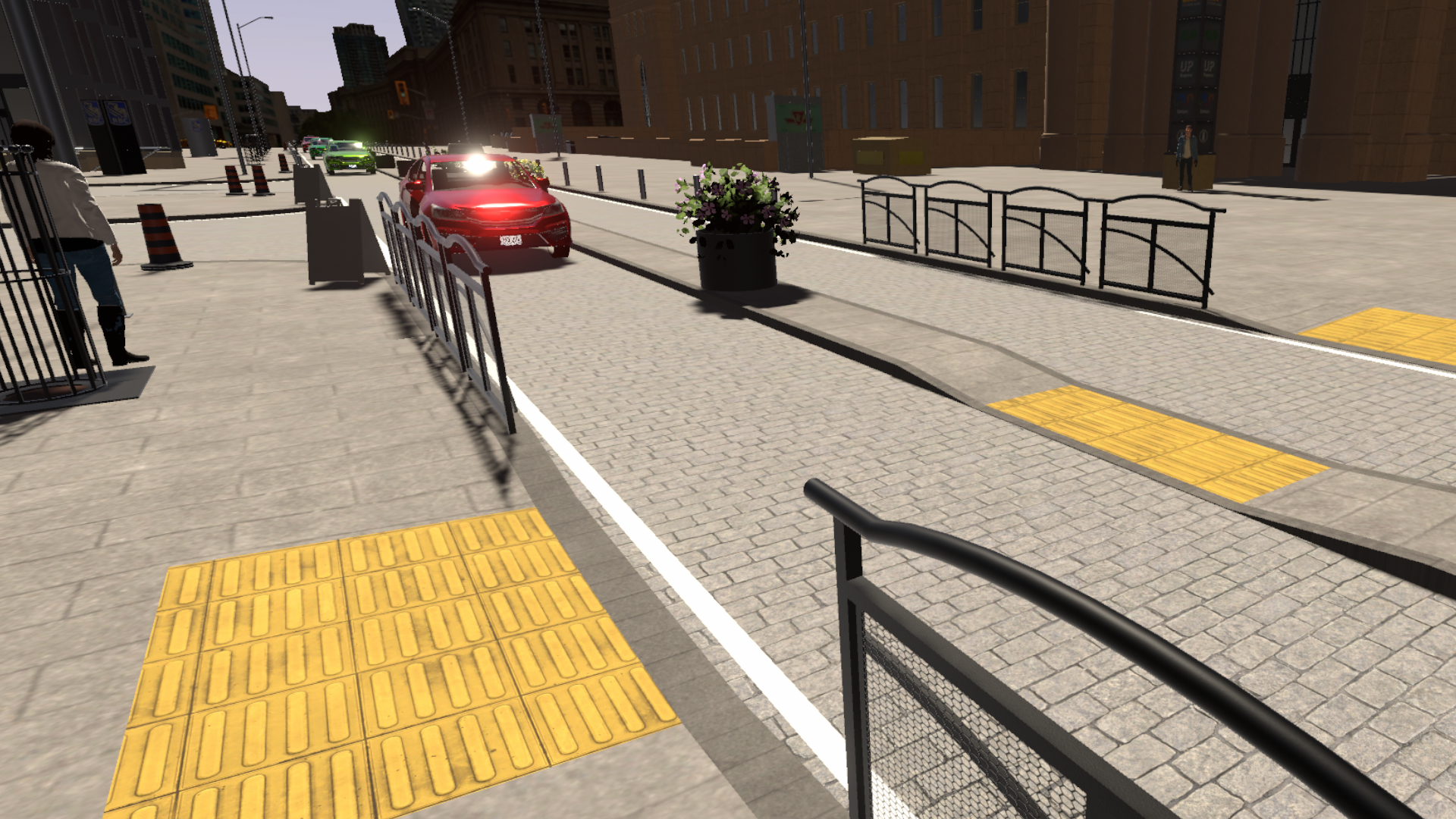}
        \caption{Street with median}
        \label{fig:median}
    \end{subfigure}%
    \hspace{0.02\textwidth}
    \begin{subfigure}{0.47\textwidth}
        \includegraphics[width=\textwidth]{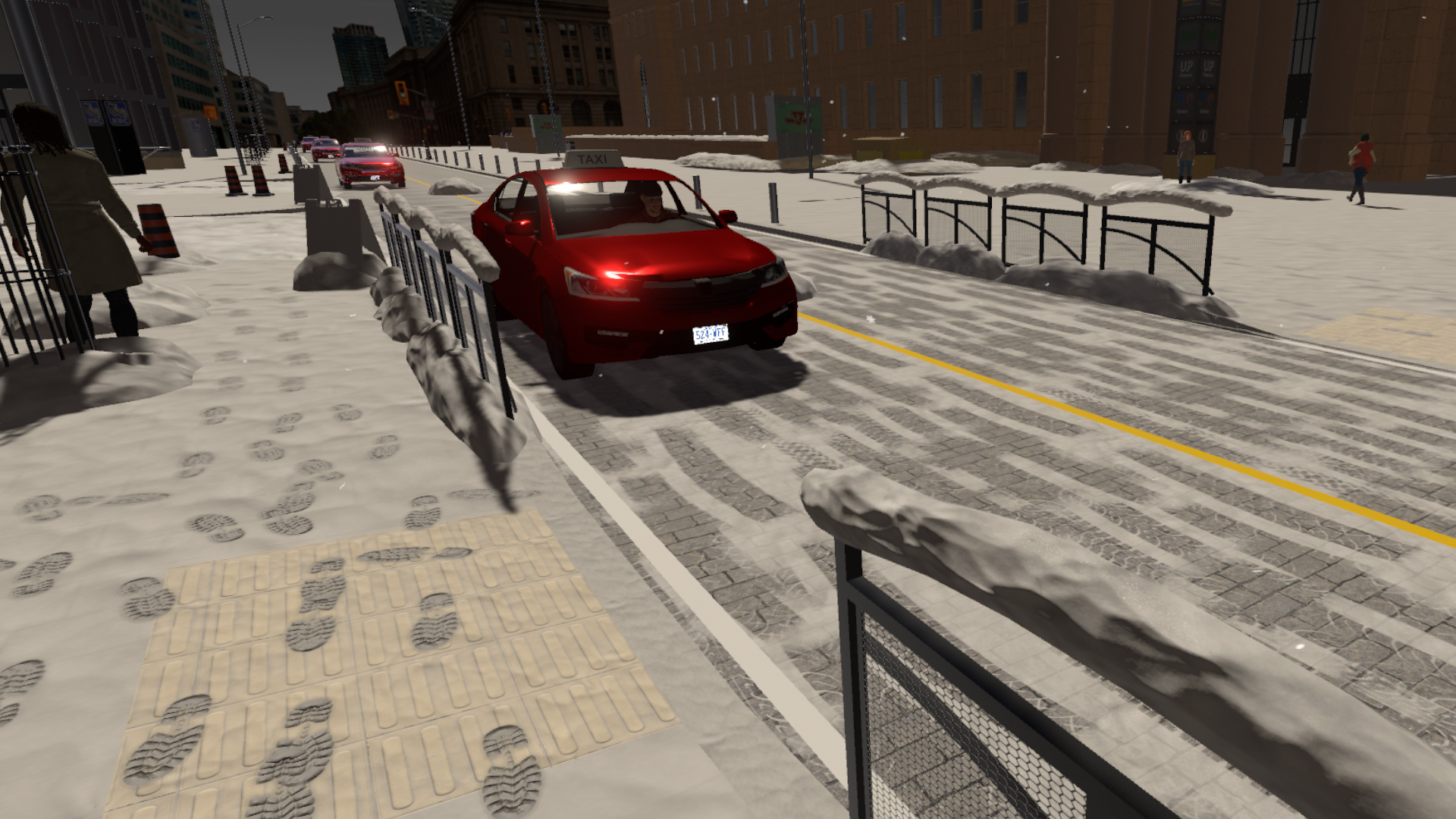}
        \caption{Snowy weather}
        \label{fig:snow}
    \end{subfigure}
\caption{A view of the simulated environments in VR}
\label{fig:simulated_environment}
\end{figure}

\begin{figure}[!ht] 
    \centering
    \includegraphics[width={0.9\linewidth}]{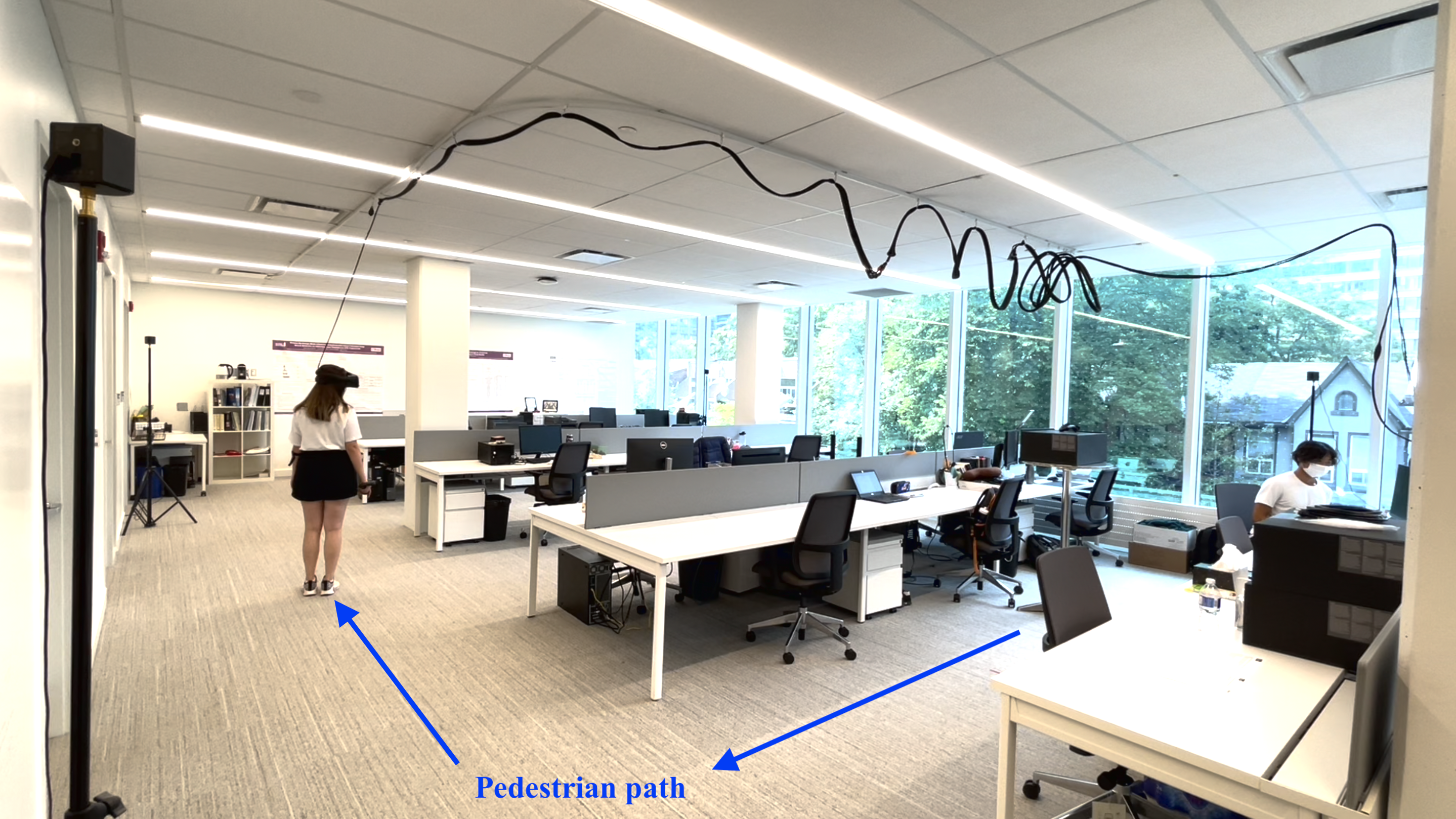}
    \caption{Walking environment of the experiment}
    \label{fig:experiment_setup}
\end{figure}

Several data collection methods and experimental tools have been applied to study pedestrian behaviour. VR is becoming an established research tool that has been validated in the context of studying the interactions between vehicles and pedestrians \citep{fengDataCollectionMethods2021, farooq2018virtual}. In this study, we used VR to achieve three key objectives. First, to ensure that scenarios could be replicated consistently for various participants. Second, to ensure the safety of participants, and third, to be able to use wearable stress monitoring sensors in a controlled environment to reduce unwanted noise. Toronto's Front Street West near the Union Station was selected as the location for the experiment. The base simulation was developed in-house using the Unity gaming platform \citep{farooq2018virtual} and then outsourced to a commercial developer for features development, generalizability, and creation of a more realistic environment and avatars (simulated pedestrians). Select views of the simulated environment are shown in Figure \ref{fig:simulated_environment}. It illustrates several simulation scenarios developed in the study to compare the stress of pedestrians while crossing the street in different situations. Figure \ref{fig:av} shows an AV signalling stop sign to the pedestrian using an eHMI panel in the front and side of the vehicle, and Figure \ref{fig:agent} shows an avatar responding to the AV eHMI signal to stop by waiting to cross. Figure \ref{fig:c} shows an AV signalling go signal using its eHMI panels, and \ref{fig:avago} shows an avatar responding to the AV eHMI signal to cross the street. Figure \ref{fig:median} shows a sunny day, while Figure \ref{fig:snow} shows the light snowy conditions. They also show the presence and absence of the median. This variation allows the study to examine if such safer crossing infrastructure and adverse weather impact the stress levels of pedestrians. A Valve Index VR headset was used to project the environment for the participants. The default experiment area for such VR headsets is 5m by 5m space, which is not enough to have a one-to-one mapping of the entire street crossing. Due to the high data rate requirements of the dynamic and immersive simulation, wireless headsets were also deemed inappropriate. Thus to replicate realistic walking conditions, we expanded the experiment area to a 10m by 10m space by installing three base stations. High data rate extension cables were also used to extend the range of the VR headset in response to the enlarged experiment area. A ceiling pulley system was designed to hold the wires above the participant's head to avoid any discomfort while wearing the headset and the possibility of tripping. Figure \ref{fig:experiment_setup} shows the experiment setup in the laboratory. Further technical details of the VR setup can be found in \ref{techdeatilsVR}. 

Electrodermal activity is a bio-signal for stress, which can be measured by various wearable sensors \citep{klimek2023wearables}. The Shimmer3 GSR+ sensor was used to monitor the EDA of each participant when walking in the VR environment. It consists of a small plastic box containing the hardware and electrodes that are attached to the skin. The unit was fastened to the non-dominant arm of the participants with two electrodes connected to their palms to read the EDA. EDA data was sampled at 100 Hz. \cite{han2020objective} reported that the Shimmer3 GSR+ sensor is 94.55\% accurate in detecting stress.

\subsection{Experiment Design and Protocol}
The experimental design was executed with consideration of all research variables and practical factors, including the maximum VR immersion duration and the overall experiment duration. Out of 1,152 possible combinations of the selected variables and their levels, 300 scenarios were selected based on their importance and relevance (See \ref{expdetails} for details). A total of 12 different sessions were randomly selected from these 300 scenarios for each participant to go through. The participant went through each scenario out of the 12 selected scenarios twice in random order. The initial combinations were tested with pilot participants to evaluate the performance of the VR experiment and observe the participant's reactions to different scenarios. All variable levels were incorporated in the final experimental design randomly, having avatar behaviour levels being repeated equally for each participant. 


A large part of the data collection was executed during the COVID-19 pandemic. Toronto Metropolitan University’s Research Ethics Board issued a Safe Scholarly, Research and Creative (SRC) Human Participant Plan for the ongoing research activities involving human subjects, which was fully observed during this research. Researchers and participants used to wear masks all the time and experiment equipment and frequently touched areas were sanitized after each participant. Along with the experiment consent forms, COVID-19 consent forms were signed to ensure that the participants were aware of the possible consequences of participating in this study. This consent form was prepared in line with the recommendations from Canadian public health authorities,  outlining additional precautions. It mentioned the potential threats of being infected by COVID-19 and the importance of being fully vaccinated, as one of the participation criteria. It guided participants to fill out the Ontario COVID-19 self-assessment tool prior to their visit to identify any symptoms. This supplementary consent form informed the participants that all research team members will follow the University’s Principles and Guidance for the Limited Resumption of Critical Human Participant and/or Field Scholarly, Research and Creative (SRC) Activity. In addition, participants acknowledged that they are not among the COVID-19 ``high risk'' category and that they had not travelled outside of Canada in the past 14 days.

Upon arrival at the experiment venue, participants were greeted. If they had not completed the consent forms online, they were asked to read and sign the experiment consent form, which contained information about the experiment, general participation criteria, and compensation. They were then provided with a COVID-19 consent form to review and sign. In parallel, an online survey was prepared to gather sociodemographic information from the participants. Once the consent forms and survey were completed, participants were briefed on the study's purpose and the experiment's details. They were given the VR equipment and instructions on how to wear the headset and operate the controller to navigate the VR instructions.

{To ensure the reliability of the EDA data and minimize the impact of confounding factors, several controlled measures were implemented. The initial onboarding and instructions processes helped to calm participants and mitigate any stressors from previous activities. The experiments were conducted in an air-conditioned room where temperature, humidity, and noise levels were kept constant to eliminate environmental influences. Additionally, data collection occurred consistently between 5 pm and 9 pm daily, to account for time-of-day effects on participant stress levels.} 

\begin{figure}[H]
     \centering
     \begin{subfigure}[b]{0.49\textwidth}
    \includegraphics[width=0.925\linewidth]{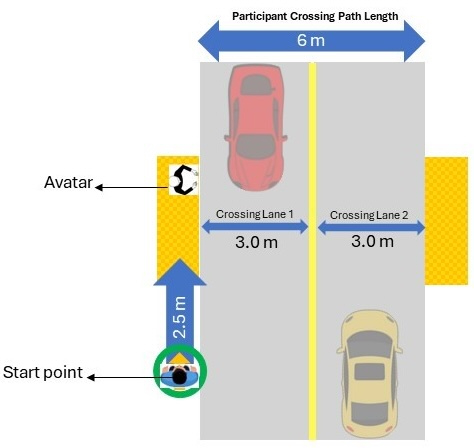}
    \caption{Without median}
    \label{fig:without_median}
     \end{subfigure}
     \begin{subfigure}[b]{0.495\textwidth}
    \hspace{-0.5cm}
    \includegraphics[width=1.075\linewidth]{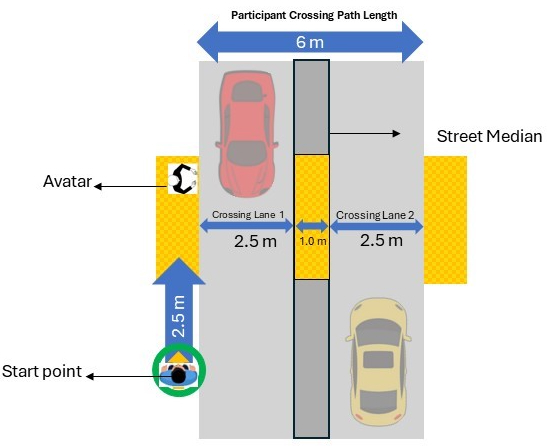}
    \caption{With median}
    \label{fig:with_median}
     \end{subfigure}
        \caption{Pedestrian crossing design}
        \label{fig:road_design}
\end{figure}

In VR, participants were placed on a sidewalk and  instructed to go to the tactile paving and then find a safe and suitable moment to cross the street. Participants were given a 60-second time limit to complete the task, and if they did not finish within that time frame, the session would be terminated and the next session would be automatically loaded. To prevent participants from experiencing discomfort, the screen was programmed to black out just before the moment of impact in the event of a vehicle accident and a message was shown to the participants indicating that they had an accident. Throughout the immersion period, a researcher accompanied the participant throughout the immersion time to ensure their safety. Figure \ref{fig:road_design} shows the details of the experimental task performed by the participant.

The participants started at the lowest point of the L-shaped course at the bottom left and progressed to the highest point at the top of the L-shaped course. The crossing contained two 3.0m wide single-lane roads; opposing traffic was on either side, requiring the participants to wait for the cars to pass before crossing the 6.0m long path. Several scenarios were compared in different road conditions, from vehicles moving in the same lane to moving in opposite directions (Figure \ref{fig:without_median}). All participants took the same simulated route back to the same point, reacted, and indicated if they felt traffic was light, normal, or heavy. The data was collected as the participants walked back to the original position where the simulation took place. This made the data collection process flow smoothly and did not distract the participants during the simulation.

Figure \ref{fig:with_median} uses the street median as another variable in the pedestrian crossing simulation. The street median is 1.0m wide, where individuals who are in the process of crossing can find a temporary safe area to wait between the two lanes of vehicles. Each lane is 2.5m wide, with a 6.0m long crossing path. Studying crossing alongside the median will help understand the influence of the street median on pedestrian stress levels under different traffic conditions within the simulated environment.

\subsection{Participants Recruitment and Features}
\label{sec:participants}

{Overall, 171 individuals participated in the experiment to ensure that all our Linear Mixed Models (LMM) have data from at least 75 individuals. Based on the approximate power analysis suggested for LMM in \cite{liu1997sample}, the minimum respondent size required was 40 individuals ($\alpha = 0.05$, $\beta = 0.2$).} The majority of participants were recruited by a panel provider. The rest of the participants were invited by developing advertisement campaigns on Instagram and TikTok. The participation criteria were as follows. 

\begin{itemize}[itemsep=-1mm]
    \item Being fully vaccinated or have received an approved exemption from the university in accordance with Toronto Metropolitan University’s COVID-19 vaccination protocols.
    \item Aged between 18 and 65 years old.
    \item Not having any head, eye, brain, or heart illness.
    \item Never had a feeling of nausea or fatigue when playing video games.
    \item To live in the Greater Toronto Area.
\end{itemize}

The dataset was processed in several steps to deal with non-existing or failed data. There were 8 cases in which the VR experiment application failed in the middle of the experiment, the tracking sensors lost track of the headset, or the scenarios were not recorded, which resulted in the termination of the experiment. The EDA data was not collected for 46 participants---in a few cases the GSR sensor failed to collect data due to Bluetooth connection failure or loose electrode connection. In multiple cases the collected EDA data was not useful---it has been found that approximately 10\% of
participants in any experiments are estimated to be non-responders (hypo-responsive) in terms of their EDA \citep{Braithwaite_Psychophysiology_2013}. The higher failure rate of the GSR sensor data in this study is rooted in the fact that the experiment setup was ambulatory so there were more chances for the electrode to get disconnected from skin.

\begin{figure}[!ht] 
    \centering
    \includegraphics[width={0.5\linewidth}]{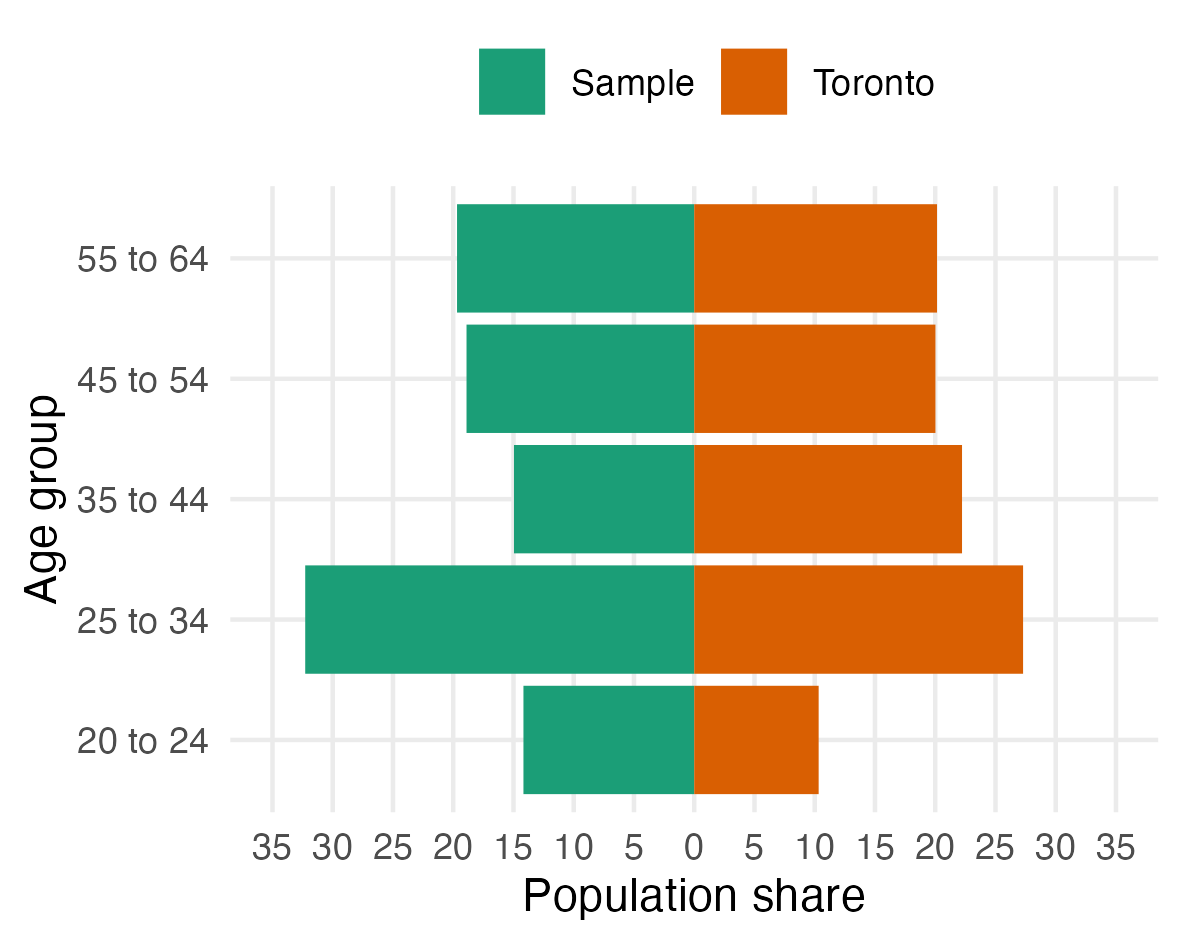}
    \caption{Age pyramid of the sample and Toronto}
    \label{fig:age_pyramid}
\end{figure}

Out of 171 participants, 37 individuals did not provide age and/or gender information. Table \ref{tab:socio} summarizes the sample characteristics. Overall, 43\% of the participants were female and 57\% were male. Our sample has approximately 8\% more males when compared to the distribution in the population of Toronto. Figure \ref{fig:age_pyramid} shows the age distribution of the sample compared to the population of Toronto in 2021 {\citep{statisticscanadaFocusGeographySeries2022}}. We observe a reasonable fit between the two distributions.

\begin{table}[htbp]
  \centering
  \caption{Sample distribution (\%)}
  \footnotesize
      \begin{tabular}{cccc}
      \toprule
      Age group & Female (43) & Male (57) & Total (100)\\
      \hline
      \hline
      18 to 24 & 6     &  8     & 14 \\
       25 to 34 &  16    &16   & 32 \\
       35 to 44 & 6     & 9     & 15 \\
       45 to 54 & 7     &       12    &     19 \\
       55 to 65 & 8     &   12    &   20 \\
      
      \bottomrule
      \end{tabular}%
  \label{tab:socio}%
\end{table}%

\section{Methods for Data Analysis}
\label{methods}
The designed VR experiment generated high-resolution, complex, and multimodal data on the participants. The resulting dataset included a) the trajectories of the participant and all the virtual objects, including vehicles and avatars at 10Hz, b) the EDA values from the GSR sensor at 100Hz, c) the audio recording of the participant while performing the task, d) recording of the first-person-view of the resulting VR simulation, and e) the associated details of the scenario the participant was in. Three complementary methods were identified to develop a systematic analysis of the collected data. Continuous decomposition analysis was adapted for data engineering and investigating trends. Systematic contextual annotation of the processed data was performed to understand the stimuli associated with the stress related arousals. In the end mixed linear model was used to quantify the effects of various stimuli.

\subsection{Continuous Decomposition of GSR Sensor Output}
In the context of emotions (fear, stress, happiness, etc.), arousal refers to a state of physiological activation in response to a stimulus. In the case of increased stress levels, it can lead to changes in the body such as an increase in sweating or heart rate. This response can vary depending on the individual and the situation and is a complex interplay between physiological state, emotions, and cognitive processes \citep{dawsonElectrodermalSystem2016}. The increase in the level of sweat increases the conductivity at the surface of the skin which can be captured by an GSR sensor. The recorded GSR signal is composed of two components:
\textit{a) Tonic Component (skin conductance level)} refers to the baseline activity level of the skin conductance. It represents the slowly varying, relatively stable background activity of the sweat glands, which is influenced by factors such as arousal level, and stress. The tonic component is often used as an indicator of the overall sympathetic nervous system arousal.
\textit{b) Phasic Component (skin conductance response)} refers to the rapid fluctuations or responses that occur on top of the tonic activity. These fluctuations are often associated with sudden changes in arousal or emotional responses to stimuli. Phasic responses can be elicited by various stimuli, such as auditory, visual, or tactile stimuli, and they reflect the transient changes in sweat gland activity induced by these stimuli \citep{Boucsein_2012}. Whenever there is an emotional reaction to an external stimulus (e.g., a vehicle approaching a pedestrian at high speed), the phasic or rapid component of the GSR signal increases its amplitude temporarily, which is known as skin conductance response (SCR).

Considering the purpose of this study, the SCR associated with the phasic component was of interest, which had to be isolated from the tonic component. Ledalab package has been widely used for this purpose and it was adopted for this study as well. One of the benefits of Ledalab is that it is run locally in Matlab and preserves data privacy. Other solutions offer web-based solutions that require uploading the data online. Details of the continuous decomposition algorithm used in this package is explained in \citet{Benedek_JournalofNeuroscienceMethods_2010}. Considering the ambulatory nature of the experiment, a threshold of 0.1 was selected for SCR detection \citep{petrescuIntegratingBiosignalsMeasurement2020}. The data were downsampled to 10 Hz to speedup the analysis. Due to the long and non-stationary behaviour of the signals, a Gaussian filter was applied with window size equal to 30---three times the downsampled data collection frequency of 10 Hz---to remove the small shakes of the signal, which largely corresponded to low-pass filtering of the data \citep{nazemiUnravellingBicyclistsPerceived2020a}. 

\begin{figure}[!ht] 
    \centering
    \includegraphics[width={0.75\linewidth}]{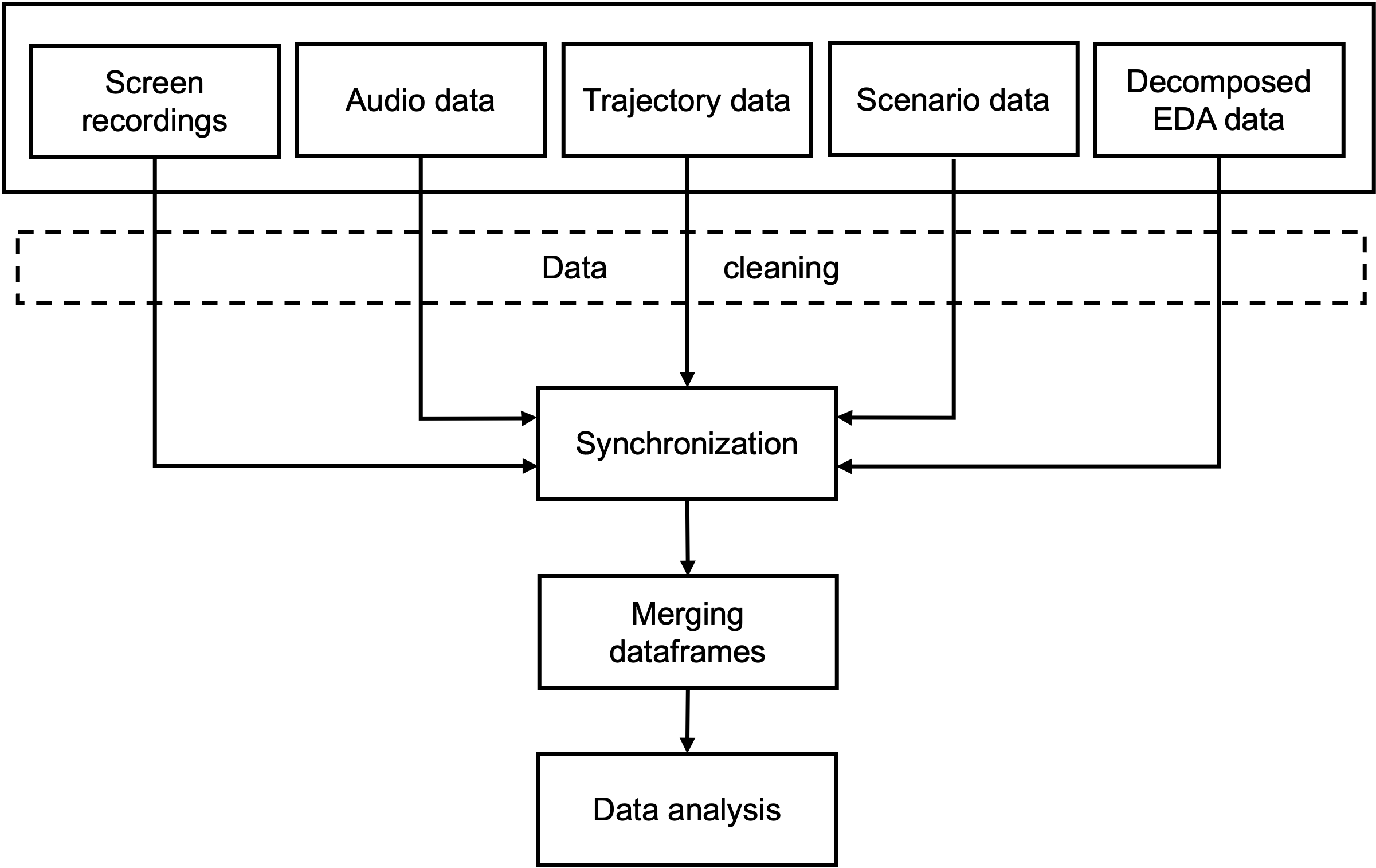}
    \caption{Data engineering workflow for the multimodal experiment outputs}
    \label{fig:workflow}
\end{figure}

The raw GSR sensor outputs were visually inspected prior to performing the above-mentioned analysis to eliminate failed readings and remove artifacts, i.e., sharp angled peaks due to movement, from the signals. The data were cleaned one more time after the above-mentioned analysis, when no phasic activity was observed. Furthermore, it was essential to synchronize the VR data and the physiological sensor data. Both times were converted into Epoch times to facilitate this process. The data engineering procedure is illustrated in Figure \ref{fig:workflow}.

Several transformation methods such as range correction, transformation into standard values, and autonomic lability scores have been suggested for SCR amplitude, due to inter-individual variations in response range, prior to performing any statistical analysis \citep{Ben-Shakhar_Psychophysiology_1985}. Transformation into standard values overcome the problem of determining individual electrodermal response maxima being a prerequisite for range
correction method \citep{Boucsein_2012}. This approach was selected for this study, which considers each particular individual mean and standard deviation of SCRs. Standardized SCRs are calculated as follows. 

\begin{equation} \label{eq:z_score}
z_{ik} = \frac{SCR_{ik}-\overline{SCR_{i}}}{s_{i}}	
\end{equation} 

\noindent where \linebreak
\(z_{ik}\) = \(k\)th SCR standardized value for individual \(i\),\\
\(SCR_{ik}\) = \(k\)th raw SCR score for individual \(i\),\\
\(\overline{SCR_{i}}\) = mean of all SCRs for individual \(i\), and\\
\(s_{i}\) = standard deviation of all SCRs.\\

The normally distributed z scores have an average of 0 and a standard deviation of 1. It is common to transform z scores to T scores with a mean of 50 and standard deviation of 10 to drop out the minus signs and freely perform a variety of statistical tests \citep{Boucsein_2012}.

\begin{equation} \label{eq:T_score}
T_{ik} = 50 + 10z_{ik}
\end{equation} 

\noindent where \linebreak
\(T_{ik}\) = \(k\)th SCR T score for individual \(i\), and\\
\(z_{ik}\) = \(k\)th SCR standardiszed value for individual \(i\).\\

\subsection{Contextual Annotation}

\begin{figure}[!ht] 
    \centering
    \includegraphics[width={0.8\linewidth}]{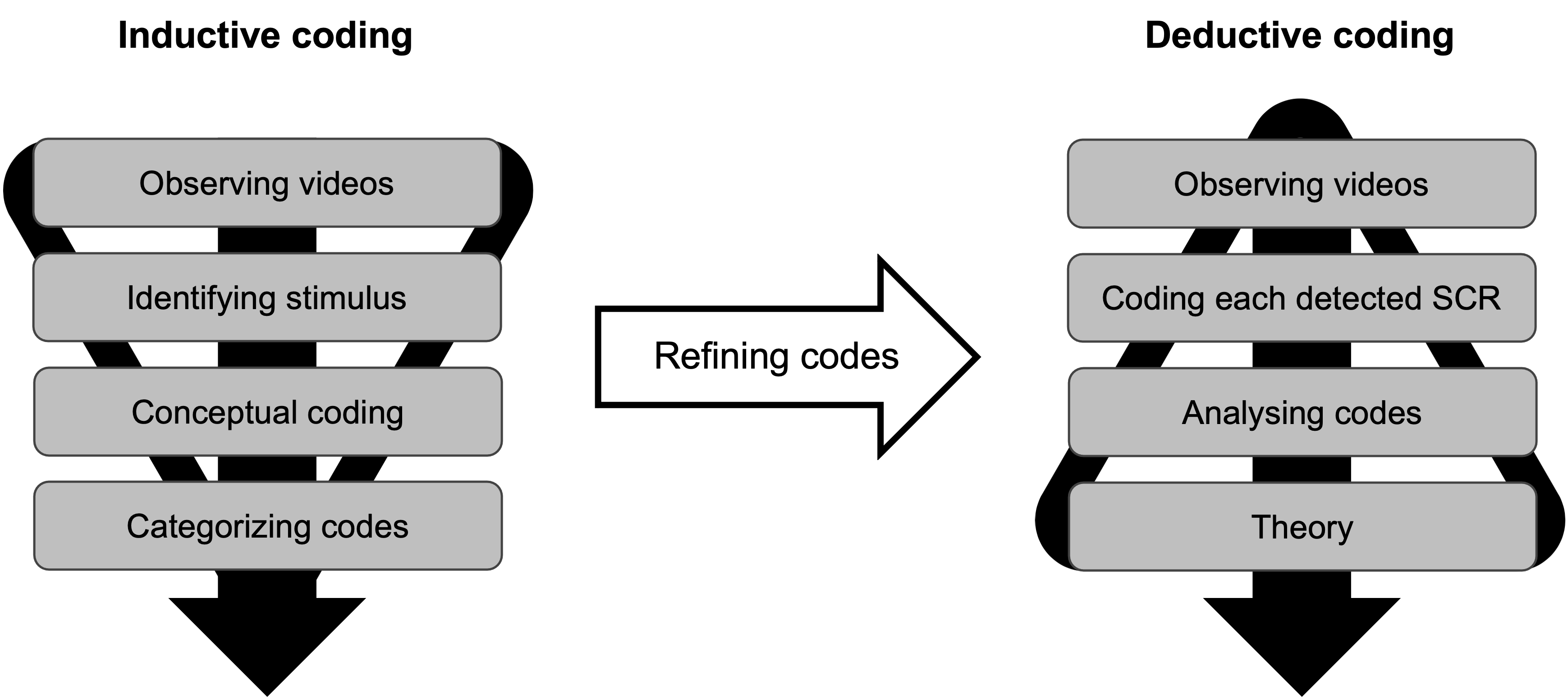}
    \caption{Annotation process diagram}
    \label{fig:annotation_diagram}
\end{figure}

While mathematical analysis of the EDA offers a strong quantitative approach to measure the arousal level of participants at any instance, it lacks the reason behind each arousal. Therefore, a qualitative coding process was performed to explore the stimuli causing arousals and identify the stress associated with crossing the street. Coding in this context refers to annotating or labelling excerpts and organizing them according to their content. \textit{Deductive coding} is a top-down approach when a researcher has a predefined set of labels to be applied to the data. \textit{Inductive coding}, on the other hand, is a bottom-up approach and includes deriving and developing different labels from the data in an exploratory manner without trying to fit into a pre-existing coding frame, or the researcher's analytic preconceptions \citep{jamesonCriticalThinkingInductive2020}. Given that the study was exploratory, i.e., investigating the data to discover patterns and to test hypotheses, a hybrid deductive and inductive approach was adopted, as shown in Figure \ref{fig:annotation_diagram} \citep{feredayDemonstratingRigorUsing2006, proudfootInductiveDeductiveHybrid2023}. The computer monitor showed the participant's view when performing the experimental task in VR. The screen was recorded for each participant while participant's online EDA level was shown simultaneously in the corner of the screen. At the same time, the participant's voice was recorded using the microphone on the VR headset. These were the input data to perform the contextual analysis. The inductive analysis involved observing a limited number of videos with the aim to identify the stimuli causing each arousal. Each potential stimulus was coded with a relevant label and the codes were categorized to identify patterns and finalize a set of labels. Subsequently, deductive coding involved observing and coding the rest of the videos using the finalized set of labels. Eventually, the codes were analyzed for hypothesis testing and deriving patterns.

\subsection{Linear Mixed Model with Panel Effects}

During the experiment, each participant completed the tasks multiple times, leading to repeated measurements of the dependent variable within the data of the same individual. This results in a correlation among response variables for each participant. Therefore, the linear mixed model (LMM) was used as the statistical tool to account for the resulting panel effects \citep{westLinearMixedModels2022}. LMM is considered a good alternative to analysis of variance in such cases. Often, ANOVA assumes the independence of observations within each group, however, this assumption may not hold in non-independent data. LMM is composed of a fixed effect component and a random effect component. Fixed effects are the parameters associated with the independent variables that are consistent across the entire population or dataset. These effects are the primary interest in most regression analyses and represent the average influence of the predictor variables. Random effects are the parameters associated with the independent variables that vary across individuals or groups of individuals that capture the variability in the effect across different participants. They allow the intercepts and/or slopes to vary for each individual or group.
The mathematical representation of the linear mixed model is presented as follows.

\begin{equation}
y = X\beta + Zu + \varepsilon
\end{equation}

In this equation, \(y\) is a dependent variable vector of size \(N \times 1\). \(X\) is a design matrix of size \(N \times p\), where \(p\) is the number of independent variables. \(N\) is the total number of observations in the dataset used for the model, calculated as:

\begin{equation}
N = \sum_{i=1}^{q} n_i
\end{equation}

Here, \(n_i\) is the number of observations for participant \(i\), and \(q\) is the total number of participants in the dataset. \(\beta\) is a vector of size \(p \times 1\), representing the fixed effects regression parameters. \(Z\) is a matrix of size \(N \times q\), representing \(q\) random effects. \(u\) is a vector of size \(q \times 1\), representing the random effects, and is normally distributed; \(u \sim N(0, G)\) where \(G\) is the variance-covariance matrix of the random effects. \(\varepsilon\) is a vector of size \(N \times 1\), representing the residuals.

\section{Results and Discussion}
\label{sec:Results}

Three complementary analyses were performed. First, the arousal levels related to different environmental, vehicle technology, social influence, traffic flow, road geometry, and sociodemographic variables were studied using LMM, without any spatio-temporal context. Subsequently, arousal levels were examined at different locations when performing the experimental task. Finally, the detected skin conductance responses (SCRs) were labelled based on the underlying stimuli and the stress associated with crossing was quantified. For each stage of the analysis, VR, EDA, video, and audio data were processed based on the type of analysis and its data requirements, resulting in different numbers of observations for each analysis.
The SCR T scores were considered as a dependent variable indicating the arousal level of participants at any instance while performing the experimental task.

\subsection{Effects of General Factors}

The variables identified in Table \ref{tab:tabexperimentfactors} combined with gender and age group were considered as independent variables to examine their influence on each of the SCR T scores. Table \ref{taballregression} presents the overall fixed and random effect parameters. Overall, it is noted that the panel effects through random parameters capture a high degree of variation, which is most probably the result of the physical (e.g., controlling certain muscle groups in certain situations) and mental (e.g., where to look first or what sequence of actions to follow) learning as the participants went through the experiment scenarios. The results revealed that having a median in the middle of a street significantly reduced the crossing stress level. A median serves as a refuge island for crossing pedestrians and would divide the crossing task into two steps, and therefore, participants incurred less stress \citep{dumbaughMostVulnerableUser2024}. The variation due to the panel effect is also relatively low. It is probably due to the fact that as the participants were going through various scenarios they learned to take advantage of the median, resulting in an average negative effect of the median on stress. Snowy weather was found to significantly reduce the incurred crossing stress. While this result seems to be counterintuitive, it can be because the pedestrians perceived drivers to be more cautious with pedestrians and driving at lower speeds in adverse weather conditions. Our analysis of the data also revealed that in snowy conditions the participants waited much longer to find the right gap and then start the crossing. This cautious strategy may have resulted in a much more confident and calmer crossing with lower stress levels. Furthermore, snowy sessions provided a brighter environment in VR with objects more clearly visible. The participants may also have felt more visible to the drivers when they were crossing the street. Therefore, they felt more relaxed compared to other weather conditions. The rainy weather resulted in an increase in the stress level, which can be due to the associated decrease in visibility and tyre traction of the vehicles. Note that the underlying VR simulation implemented the physics behind the tyre traction in clear, rainy, and light snow conditions.


The age group 18 to 24 was found to be least stressed when crossing the road, while the age group 55 to 65 experienced significantly higher stress levels. This is consistent with the results by \cite{farooq2022} who reported that the pedestrians over 50 and those in the 40–49 age bracket considered the road condition to be more stressful, with the opposite effect reported for the 30-39 age group. Such findings suggest that age significantly affects how subjects respond to the variables in the urban environment, such as experience, cognitive function, and state of their mobility, across different age groups. So, it is essential to introduce these features as they meet many other needs that most of the demographics require with the development of urban designs. This could include increased accessible walking signals, lighting, crossing times, and pedestrian infrastructure. For this purpose, adaptation and accommodation of urban spaces to the needs of all age groups are vital to achieving safer and more inclusive cities. 

There were no statistically significant fixed effects in the arousal levels when controlling for vehicle type, traffic flow, time of day, and gender. The results indicate that pedestrians did not get stressed when they had to deal with the AVs when compared to the normal vehicles. This finding is in line with the previous field research which concluded that implicit communication cues are more likely to be used by pedestrians while crossing the road, compared to explicit communication cues from drivers \citep{leeRoadUsersRarely2021, mooreCaseImplicitExternal2019}. However, it is noted that the associated random effect parameters were statistically significant and with a high magnitude, which again can be due to mental and physical learning as the participants went through different scenarios.

\subsection{Effects of Social Influence}
During the experiments, it was observed that the avatar's actions only influenced participant arousal levels either when they were on the yellow tactile paving waiting to cross or when they started crossing in the first lane shortly after the avatar. Therefore, only these locations were chosen to examine variations in the detected arousal levels related to the avatar's actions. The difference in the arousal levels relative to each avatar action was modelled using the LMM reported in Table \ref{tabavatarregression}. The parameter for ``No avatar" was set as a reference. Having a standing avatar waiting to cross the street had a statistically insignificant effect on pedestrian stress. In the case of a conservative avatar, the pedestrian stress level increased when compared to no avatar. The increase in stress level was even more in the case of an adventurous avatar. In several cases, participants even interjected during the experiment, expressing shock at the risky behaviour of the adventurous avatar. The relatively significant random effect parameters for all three cases suggest that it took the participants a bit of learning to understand the behaviour of the avatar and get used to it. These results point to the existence of a varying level of social conformity in the crossing behaviour of the pedestrian. When the participants saw an avatar exhibiting certain behaviour, they also wanted to replicate it and as a result became more stressed.

\begin{table}[htbp]
  \centering
  \footnotesize
  \caption{Model results for SCR T scores across different factors}
    \begin{tabular}{lrrrr}
    \toprule
    Independent Variable & \multicolumn{4}{c}{LMM Estimates} \\
    \midrule
          & Fixed ($\beta$) & t-value & Random ($\sigma$)& t-value \\
\cmidrule{2-5}    (Intercept) & 54.08 & 28.66 & -- & -- \\
          &       &       &       &  \\
    \textbf{Vehicle type} &       &       &       &  \\
    Normal vehicles & --    & --    & --    & -- \\
    Autonomous taxis with roof sign & 0.66 & 0.67  &16.06 & 4.01 \\
    Autonomous vehicles with eHMI & -0.78 & -0.75 & 20.51 & 4.53\\
          &       &       &       &  \\
    \textbf{Traffic flow} &       &       &       &  \\
    Low arrival rate \& high speed & --    & --    & --    & -- \\
    High arrival rate \& low speed & 0.88  & 1.01  & 8.41 & 2.90 \\
          &       &       &       &  \\
    \textbf{Street median} &       &       &       &  \\
    No median & --    & --    & --    & -- \\
    With median & -1.24 & -1.38 & 4.95 & 2.22 \\
          &       &       &       &  \\
    \textit{Time of day} &       &       &       &  \\
    Day   & --    & --    & --    & -- \\
    Night & -0.76 & -0.76 & 23.95 & 4.89 \\
          &       &       &       &  \\
    \textbf{Weather} &       &       &       &  \\
    Sun   & --    & --    & --    & -- \\
    Rain  & 0.84  & 0.72  & 15.31 & 3.91 \\
    Snow  & -2.76 & -2.79 & 10.97 & 3.31 \\
          &       &       &       &  \\
    \textbf{Gender} &       &       &       &  \\
    Female & --    & --    & --    & -- \\
    Male  & -0.32 & -0.31 & -- & -- \\
          &       &       &       &  \\
    \textbf{Age group} &       &       &       &  \\
    18 to 24 & --    & --    & --    & -- \\
    25 to 34 & 4.41  & 2.62  & --  & --\\
    35 to 44 & 4.02  & 2.13  & -- & -- \\
    45 to 54 & 4.47  & 2.27  & --  & --\\
    55 to 65 & 7.37  & 4.08  & --  & -- \\
    \midrule
    Number of participants & 76    &       &       &  \\
    Number of observations & 1270  &       &       &  \\
    \bottomrule
    \end{tabular}%
  \label{taballregression}%
\end{table}

\begin{table}[htbp]
	\centering
	\footnotesize
	\caption{Model results for SCR T scores regressed on avatar's behaviour}
	\begin{tabular}{lrrrr}
		\toprule
		Independent Variable & \multicolumn{4}{c}{LMM Estimates} \\
		\midrule
		& Fixed ($\beta$) & t-value & Random ($\sigma$)& t-value \\
\cmidrule{2-5}    (Intercept) & 48.35 & 188.46 & -- & -- \\
          &       &       &       &  \\
    \textbf{Avatar behaviour} &       &       &       &  \\
    No avatar & --    & --    & --    & -- \\
    Standing avatar & -0.06 & -0.19 & 2.35 & 1.53 \\
    Conservative avatar & 0.96  & 3.08  & 3.92  & 1.98 \\
    Adventurous avatar & 2.89  & 4.80  & 17.87  & 4.23 \\
    \midrule
    Number of participants & 108   &       &       &  \\
    Number of observations & 7847  &       &       &  \\
    \bottomrule
    \end{tabular}%
  \label{tabavatarregression}%
\end{table}%

\subsection{Locational Analysis of Stress}
The walking manoeuvre of participants depicted in Figure \ref{fig:experiment_setup} was divided spatially into four different segments to perform the analysis (refer to Figure \ref{fig:road_design}), a. Sidewalk, b. Waiting to cross (yellow tactile paving), c. Crossing lane 1, and d. Crossing lane 2. Segments ``Crossing lane 1" and ``Crossing lane 2" were areas where the crossing task was performed and therefore, they were selected to investigate potential variations in detected arousal levels. 

The differences in the detected SCRs for each segment are further explored by means of a mixed-effects model on the panel data (see Table \ref{tab:t_segments_regression}). Here the sidewalk segment is considered the reference segment. The results show that a participant's stress level was found to be significantly higher when on the street, compared to the time when they were on the sidewalk. This model was able to quantify that being on the road increases the stress level of pedestrians by approximately 8 units on the scale of SCR T scores, compared to the time when they are on the sidewalk. \cite{li2013} stated that sidewalks with roadside plants produce a more pleasant environment, which was proven not only subjectively but also objectively. Pedestrians express greater comfort and less stress when walking along tree-lined sidewalks than on bare streets. These physiological measurements solidify claims that pedestrians feel more comfortable with less stress if they walk on tree-lined sidewalks instead of walking on a naked street. More so, this also shows how much environmental factors affect pedestrian stress. 

Participants were evaluating the traffic flow when they were waiting to cross and the process of checking the incoming traffic, finding a suitable gap, and sometimes hesitation to cross increased arousal levels. Therefore, the arousal level of participants found to be significantly higher on the ``Waiting to cross segment", compared to the sidewalk segment. Since the traffic was initially only coming from one side of the road and after a while from the opposite direction, it happened that the participants did not expect vehicles when crossing the second lane, which sometimes resulted in having accidents and large arousal levels due to shocking. Participants were also experiencing high arousal levels when standing insecurely in the middle of the roads with no median, checking the traffic stream that was coming from the opposite direction. As reported in Table \ref{taballregression}, street medians decrease a pedestrian's exposure time to traffic, which would reduce stress since the person can concentrate on crossing one direction of traffic at a time rather than the whole roadway at once. \cite{stipancic2019} reported that the presence of a median is observed to reduce pedestrian injuries significantly by 24\%. \cite{batomen2023} also reported a declining trend of collisions and injuries over time, with a median incidence rate ratio for collisions of 0.88 (95\% credible interval (CrI): 0.63, 1.23). This means that the rate of collisions was 12\% lower over the study period, indicating the effectiveness of certain conditions like traffic-calming measures. In addition, the existence of medians reinforces what pedestrians view as safe to cross streets, hence, increase perceived safety for pedestrians \cite{pedestrian_safety2019}. Medians also serve as traffic calming measure, a study by \cite{martinelli2022} reported that installing median barriers results in reduction in vehicles speed through its effect on driver behaviour providing a greater perceived safety. Medians lower the speed of vehicles, significantly decreasing the probability and severity of crashes, and reducing pedestrian stress and increasing their safety.  

The random effect parameters associated with different segments exhibit some interesting trends. The parameter is statistically insignificant for waiting to cross segment, while significant for crossing lanes 1 and 2 segments. The magnitude of the significant parameters is relatively high too. When the participant is standing on the yellow tactile paving and waiting to cross, only the memory and mental learning accumulated over the scenario may come into play. However, when the participant is actually crossing on the other two segments, physical learning and muscle memory also become a factor. This could be one of the possible reasons for higher variation in the random parameter for crossing segments.

\begin{table}[htbp]
	\centering
	\footnotesize
	\caption{Model results for SCR T scores across different segment of the road and sidewalk}
	\begin{tabular}{lrrrr}
		\toprule
		Independent Variable & \multicolumn{4}{c}{LMM Estimates} \\
		\midrule
		& Fixed ($\beta$) & t-value & Random ($\sigma$)& t-value \\
\cmidrule{2-5}    (Intercept) & 47.48 & 250.26 & -- & -- \\
    \textbf{Segment} &       &       &       &  \\
    Sidewalk & --    & --    & --    & -- \\
    Waiting to cross & 0.88  & 3.85  & 0.79  & 0.89 \\
    Crossing lane 1 & 7.53  & 11.05 &  31.93 & 5.65 \\
    Crossing lane 2 & 9.25  & 15.82 & 29.44  & 5.43 \\
    \midrule
    Number of participants & 122   &       &       &  \\
    Number of observations & 12354 &       &       &  \\
    \bottomrule
    \end{tabular}%
  \label{tab:t_segments_regression}%
\end{table}%

\subsection{Effects of Stimuli on Stress}
Here we analyzed the associated events and context behind individual SRC for each pedestrian to understand the stimuli and quantify their effects.
Initially, the data for five randomly selected participants were selected. Two highly trained analysts independently coded the detected SCRs to explore different stimuli that could be associated to the arousals. The researchers first viewed the videos of those five participants and openly developed a set of labels for each detected major arousal. Coding was based on what participant had observed in VR, their behaviour, and their interjections and comments during their VR interaction. Parallel inductive coding yielded to a set of 16 labels in the first round of coding. These labels were reviewed by the researchers and they were consolidated into 13 labels. The coding process and labels were then reviewed and revised by an expert supervisor which resulted in reducing the set of labels into 10 labels after negotiation (see Table \ref{tab:codedlabel}). In the next step, the first 10 participants were coded separately and compared for agreement to ensure an objective analysis of the results. The whole dataset was then coded deductively with these 10 finalized labels. On average, it took approximately one hour to watch 12 sessions and code each arousal for each participant. Those arousals that could not be associated to any stimulus were coded as \textit{Unknown}. The list of the labels and their definitions are presented as follows. The ``Immersion" label was omitted from the analysis as it did not align with the research objectives. The complete labelling of the data for one participant required an average of two hours from a researcher. A detailed example of the complete labelling process is presented in \ref{applabellingprocess}.

\begin{table}[!ht]
    \centering

    \caption{Coded labels and their definitions}
    \footnotesize
    \begin{tabular}{p{4cm} p{11cm}}
    \toprule
    \textbf{Label} & \textbf{Definition}\\
    \hline
    Immersion & The arousal associated with being immersed in a new VR environment\\
    Avatar's action & The arousal associated with observing the avatar's action\\
    Traffic speed & The arousal associated with high traffic speed, when the participant perceives it as high and verbally expresses it\\
    Checking far-side traffic & The arousal associated with checking the far-side traffic\\
    Hesitation to cross & The arousal associated with uncertainty in making the choice of crossing, when the participant is checking for the near-side and far-side traffic\\
    Decision to cross & The arousal associated with making the decision to cross the street, usually followed by moving forward\\
    Crossing & The arousal associated with crossing the street\\
    Fear of accident & The arousal associated with the situation when the participant thinks that there is a high chance of having an accident\\
    Accident & The arousal associated with having an accident\\
    Unknown & The arousals that could not be unequivocally linked to any of the aforementioned labels \\
    \bottomrule
    \end{tabular}
    \label{tab:codedlabel}
\end{table}

Similar to the previous analyses, the differences in the mean detected SCRs for each label are further explored by means of mixed effects model on the panel data. The mean T scores of the detected SCRs are taken as the dependent variable. The labels were included as independent variables with ``Crossing" as the reference factor. The model results are presented in Table \ref{tab:t_labels_regression}. The ``Crossing" label was used as a reference. 
The stress level associated with the ``Fear of accident" and ``Accident" labels were found to be significantly larger than the other labels, and approximately 22 and 20 units more than the crossing stress, on the scale of SCR T scores. It should be noted that in certain accident scenarios, individuals did not see that a car collided with them; they only saw the dark screen and the message that informed them that they had an accident. The incurred stress associated with crossing is significantly higher than the emotional responses that occurred when the participant was on the sidewalk, i.e., avatar's action, checking the far-side traffic, hesitation to cross, and the decision to cross. The random effect parameters showed a similar trend as previous analyses, with a relatively high magnitude, which is possibly associated with the learning effects. Note that the magnitude of within-participant variance for the labels accident and fear of accident is very high. This is associated with the virtual nature of the experiment. The participants quickly get used to the virtual nature and may start feeling safer as they realize that the vehicles cannot actually harm them. Additionally, they also come to know that if an accident is imminent the simulation will automatically stop and they would not experience even the virtual accident.

\begin{landscape}
\begin{figure}[!ht]
\vspace{-1cm}
	\centering
	\begin{subfigure}{0.875\linewidth}
        \centering
		\includegraphics[width=\linewidth]{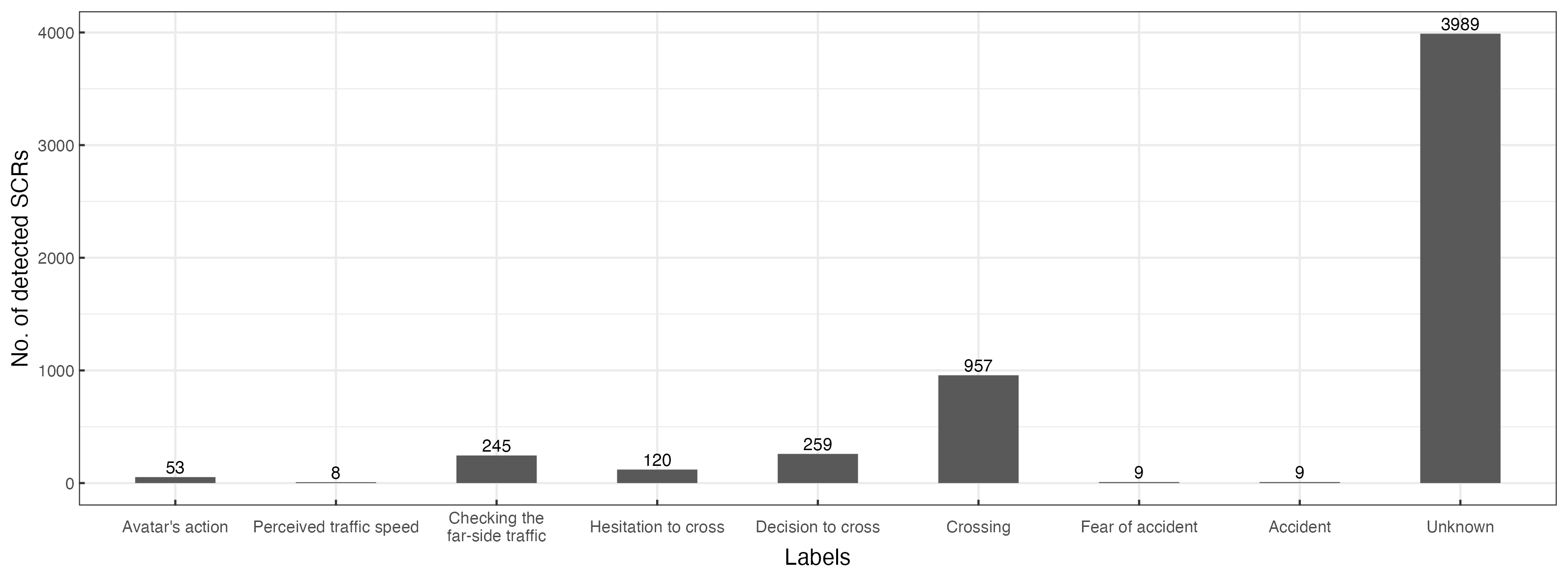}
		\caption{Number of detected SCRs for each label}
		\label{fig:subfigA}
	\end{subfigure}
	\begin{subfigure}{0.875\linewidth}
        \centering
		\includegraphics[width=\linewidth]{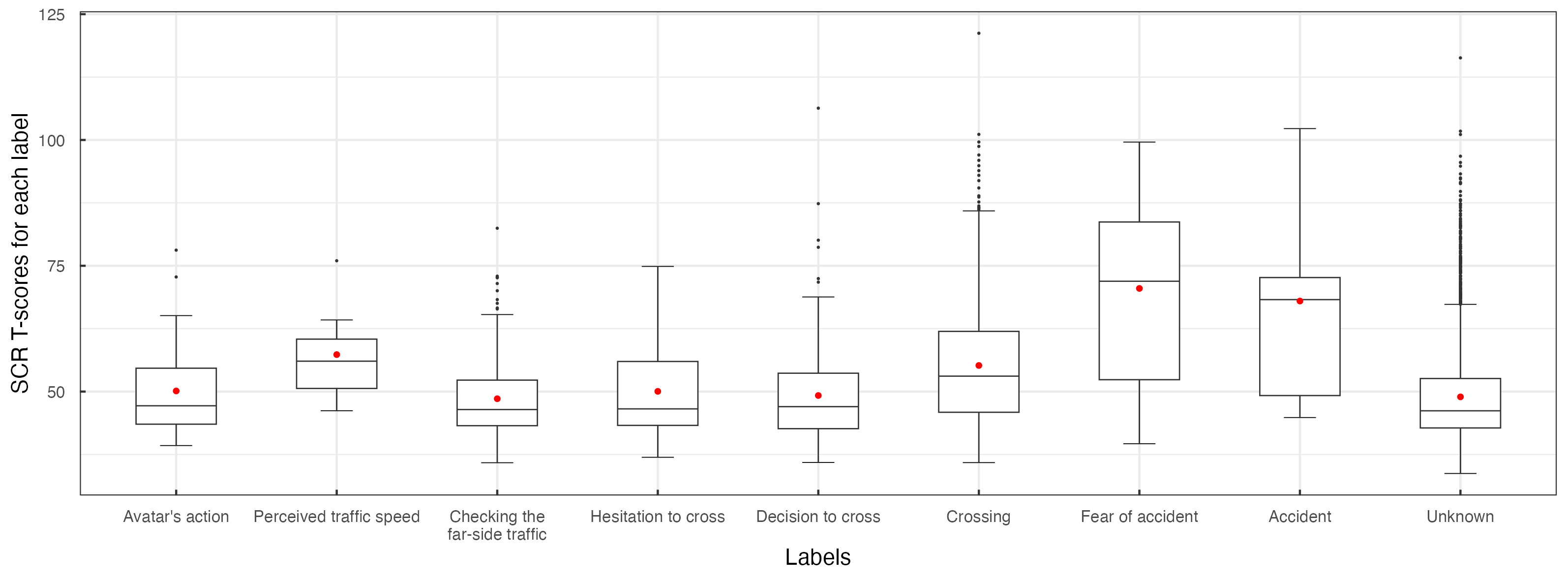}
		\caption{Box plot of SCR amplitudes for each label}
		\label{fig:subfigB}
	\end{subfigure}
	\caption{Number of detected SCRs and SCR amplitudes for each label (No. of participants = 89)}
	\label{fig:label_no_mean_SCR}
\end{figure}
\end{landscape}

Figure \ref{fig:label_no_mean_SCR} shows the frequency and the average arousal level of each label. Almost all crossings were associated with an arousal and it was fairly easy for analysts to identify such arousals. That is why ``Crossing" is the most frequently observed label. Each crossing was usually preceded by a ``Decision to cross" arousal which happens right before the participant selects a gap and crosses the street. ``Fear of accident", ``Accident", and ``Traffic speed" are the less frequently observed labels. the highest stress levels are related to ``Fear of accident", and ``Accident".

\begin{table}[htbp]
	\centering
	\footnotesize
	\caption{Model results for SCR T scores across different stimuli}
	\begin{tabular}{lrrrr}
		\toprule
		Independent Variable & \multicolumn{4}{c}{LLM Estimates} \\
		\midrule
		& Fixed ($\beta$) & t-value & Random ($\sigma$)& t-value \\
\cmidrule{2-5}    (Intercept) & 55.26 & 109.85 & -- & -- \\
    \textbf{Label} &       &       &       &  \\
    Crossing & --    & --    & --    &  \\
    Avatar's action & -5.08 & -3.27 &15.99 & 4.00 \\
    Traffic speed & 1.23  & 0.31  & 63.74& 7.98 \\
    Checking the far-side traffic & -6.82 & -8.31 & 7.52 & 2.74\\
    Hesitation to cross & -4.24 & -3.59 & 10.06 & 3.17 \\
    Decision to cross & -6.04 & -7.26 & 9.31 & 3.05 \\
    Fear of accident & 14.92 & 2.39  & 300.40  & 17.33 \\
    Accident & 3.40 & 0.56  & 108.97  & 10.44 \\
    \midrule
    Number of participants & 75    &       &       &  \\
    Number of observations & 1660  &       &       &  \\
    \bottomrule
    \end{tabular}%
  \label{tab:t_labels_regression}%
\end{table}%

\section{Policy Implications}
With regard to policy implications and design interventions, this study demonstrated how having medians or islands in the middle of two-way streets significantly reduces the crossing pressure of pedestrians. It was observed that pedestrians find medians a safe place where they can pause and divide the street crossing process into two stages. In larger blocks where mid-block crossings are common, and in more crowded streets with higher volumes of vehicles and pedestrians, it is recommended to implement traffic calming solutions, raised pedestrian crossings, or pedestrian crossovers to ensure safety of pedestrians. The proposed framework in this study is capable of quantifying the reduced stress under any of the suggested design interventions, providing a valuable tool for urban planners and policymakers to evaluate the effectiveness of these measures.  

While there have been several studies on the design of eHMI systems for AVs to prompt pedestrians when it is safe to cross in conflict areas, the results revealed that pedestrians primarily base their decisions on vehicle-based movement information such as yielding cues and gaps. Utilizing eHMIs to communicate when it is safe to cross did not significantly reduce pedestrian crossing stress when compared to conventional cars and AVs with roof signs. However, eHMI systems can enhance implicit interactions between vehicles and pedestrians by communicating their intentions to pedestrians, e.g., indicating deceleration before approaching conflict areas. In this study, we only consider a text and signs based signal. Future studies should extensively explore various other eHMI systems with both audio and visual outputs to determine the most useful communication options.

Another important aspect is raising awareness which can be implemented through pedestrian education and learning programs, particularly concerning interactions with AVs. VR setups like the one used here can be a great tool to develop mental and physical learning and increase awareness about AVs. Public campaigns can educate pedestrians on the different signals and behaviours of AVs, which can help reduce uncertainty and stress when crossing streets. These programs can be integrated into general traffic information campaigns and targeted to directly help several age groups, especially the older population, who witnessed high stress levels in the experiment. Integrating pedestrian safety measures that account for varying demographics and information regarding AVs in the general design of urban settings would greatly benefit urban planners and policy-makers in designing safer and more inclusive environments.

This study's findings highlight the need for enhanced visibility and predictability in urban environments, which would benefit pedestrians and drivers. Clearer street signs, better lighting, and the use of advanced materials to mark crosswalks and pedestrian walkways can go a long way toward such innovations. Sensors that can be embedded into smart crosswalks can ease the pedestrian and AV interactions ensuring both sides are alert while on the road. Moreover, the use of real-time traffic information can ensure the safety of pedestrians if they are able to know in advance that an intersection is unsafe or if the crossing zone is truly dangerous at that particular time. Implementing these changes effectively will help in developing a much friendlier and more secure urban environment for pedestrians.  

\section{Conclusions}
\label{sec:conclusions}

This paper reports a novel experiment implemented in a VR environment to analyze major factors influencing the stress experienced by a pedestrian when crossing an uncontrolled street. While most of the recent studies focused on different linear measures, like the time a pedestrian takes to cross and vehicle speed, this experiment collected actual EDA data from participants at stressful moments and locations during the crossing using a GSR sensor. The effects of environment, road geometry, traffic conditions, vehicle technology, social conformity, and sociodemographics on the arousal levels of the participants were explored. 

The presence of geometric features such as a median decreased the stress experienced by the pedestrian. The results showed that the behaviour of the avatars affected the levels of arousal significantly. The presence of an adventurous or conservative avatar resulted in more stress levels than an avatar standing still or no avatar at all. Looking further into the causes of arousal, there was no significant difference in terms of stress levels between different vehicle types and different AVs with the eHMI signals. However, it must be highlighted that in our initial pilots, 28\% of the participants could not clearly distinguish regular vehicles and AVs with roof signs or AVs with eHMI in a VR environment. This shows the need for carefully designed awareness campaigns and training programs that can highlight the capabilities and features of AVs, particularly the ones that can further improve the experience of pedestrians on urban roads. A high degree of physical and mental learning was observed and captured by incorporating panel effects in the developed models.

Despite the promising nature of the results, there are some limitations in this study that can be investigated in future research. It should be noted that in some cases participants performed the crossing task before fully observing the avatar's actions. Avatar's actions were triggered at a certain time in each session and controlling the participant's crossing to happen after this time was one of the limitations of this study. The influence of social conformity on the decision-making process of the participants should be studied separately in the context of the theory of planned behaviour/reasoned action. The moment a participant decided to cross the street should be compared to the relative avatar's action time and it should be considered as the response variable, while participant attitudes, subjective norms and other environmental and physiological variables can be considered as predictors. {The participants of this study involved healthy individuals who can use VR, while excluding individuals with disabilities, elderly, children, and other most vulnerable groups. Specialized experiments need to be developed that can accommodate these groups.}

The combination of VR and an GSR sensor proved to offer a powerful and safe research method for pedestrian studies. Furthermore, a big feature of this platform is the replicability of scenarios and full control over the traffic characteristics and environmental factors (e.g., time of day, weather, street design, etc.). However, there are still questions regarding the validity of the results compared to the real-world conditions, which need further investigation. Yet, it should be noted that it is not possible to control the crossing situation to isolate the stress effects in a field study. Future research work can involve incorporating kinematic variables such as vehicle motion cues and time to collision, and other variables such as wait time in studying the arousal levels. Another potential direction for future research involves integrating the arousal levels of pedestrians into choice models for their gap acceptance/rejection behaviour.

\section*{Authorship contribution statement}
\textbf{M. Nazemi:} project administration, methodology,
data collection, data analysis, original draft, writing, review, and editing the manuscript. 
\textbf{B. Farooq:} conceptualization, experiment design, methodology, project administration, investigation, resources, original draft, original draft review and editing, supervision, and funding.
\textbf{B. Rababah:} data collection, data analysis, original draft. \textbf{D. Ramos:} data collection, data analysis, software. \textbf{T. Zhao:} methodology, software

\section*{Acknowledgements}
This work is based on an Economic and Social Research Council of United Kingdom funded research project called VERONICA and is co-funded by the Canada Research Chair fund. The authors would like to express their gratitude to the current and previous LiTrans members and the project associates who generously provided their support. The authors would like to acknowledge the Centre for Urban Innovation building manager for his continuous support and for providing a dedicated space for this VR experiment throughout this extended period.

\appendix

\section{{Technical Details of VR Setup}}
\label{techdeatilsVR}
\setcounter{figure}{0}
\renewcommand{\thefigure}{A\arabic{figure}}
\setcounter{table}{0}
\renewcommand{\thetable}{A\arabic{table}}

{The environment was created using Unity, a game engine that allows developers to create simulations with VR. In order to utilize Unity with our VR headset, the Valve Index, we also required Steam VR, a VR platform developed by Valve Corporation, primarily designed to work with the Steam gaming platform. Our computer hardware's major components consists of a 13th Gen Intel(R) Core(TM) i9-13900KF with 3.00 GHz as the processor, 64 GB of RAM, and a NVIDIA Geforce RTX 4090 as the graphics card. Although we invested into a computer with powerful components that allows the experiment to run smoothly, this experiment can also be run on an average computer just as well.  In order to ensure the VR environment is running smoothly, the refresh rate in SteamVR must be calibrated. The refresh rate, measured in Hz, indicates the amount of times per second a screen updates with new display information. A high refresh rate is necessary as it counteracts motion blurring and eye strain, improving clarity and reducing motion sickness for the participants. However, a refresh rate too high is too computationally expensive to maintain, therefore the refresh rate for the experiment is set to 120 Hz. The experiment virtual agents, the vehicles and other pedestrians in the scene, were simulated as spawned gameobjects when the experiment starts, meaning the vehicles are continuously spawned from a point on the road and the pedestrians are spawned from set positions on the sidewalks. The functionality of the gameobjects are determined by the parameters we give to the vehicles and crossing agents. These parameters affect the spawned vehicle's speed, stopping distance, agent's crossing behaviour, and many other behaviours. The rendering setting used was the Universal Rendering Pipeline (URP), a reliable rendering technique with balanced performance and lighting settings. This is the default pipeline used by VR applications when initializing a VR project for Unity and works with the majority of VR applications created.}

\section{{Additional Details on Experiment Design}}
\label{expdetails}
\setcounter{figure}{0}
\renewcommand{\thefigure}{B\arabic{figure}}
\setcounter{table}{0}
\renewcommand{\thetable}{B\arabic{table}}
{We used two pilots with a total of 50 participants to finalize the design. Some key details of the process are as follows:}
\begin{itemize}
    \item {All variables and their levels listed in Table \ref{tab:tabexperimentfactors} were considered, while discretizing naturally continuous variables, such as speed.}
    \item {Impossible combinations were constrained; for example, stormy weather and clear sky cannot happen at the same time, or eHMI is only applicable for AVs, and not for normal vehicles.}
    \item {Variables were classified into 3 tiers in terms of importance for achieving the research goals:}
    \begin{itemize}
        \item {Tier 1: Agent behaviour (4 categories), AV signal (2), Vehicle share (3), Arrival rate (2)}
        \item {Tier 2: Max speed (2), Median (2)}
        \item {Tier 3: Time of day (2), Clouds \& Precipitation (3)}
    \end{itemize}
    {This resulted in 1152 unique scenarios.}
    \item {A difficulty score was calculated for each scenario according to its conditions to make sure pilot participants do not go through easy or difficult scenarios subsequently and the required attention and effort are distributed. Based on the initial testing, we devised a point system and ranked the difficulty of each variable category. The weighting is listed in Table \ref{tab:difflevel}.} 
    \begin{table}[h]
    \centering
    \footnotesize
        \caption{Difficulty points for various variables.}
    \label{tab:difflevel}
    \begin{tabular}{ll}
    \toprule
    \textbf{Variable}      & \textbf{Points of Difficulty}       \\ \hline
    AV                     & 0 pts for all levels                                \\
    Arrival rate           & Low = 0, High = +1 pts             \\
    Signal                 & 0 pts for all levels                                \\
    Agent behavior         & 0 pts for all levels                                \\
    Speed                  & Slow = 0 pts, High = +1 pts        \\
    Median                 & 0 pts for all levels                \\
    Time of Day            & Day = 0 pts, Night = +1 pts        \\ 
    Clouds                 & Clear = 0 pts, Storm = +1 pts      \\ 
    Precipitation          & Clear = 0 pts, Rain = +1 pts       \\ \bottomrule
    \end{tabular}
    \end{table}

    \item {An orientation session/scenario was considered the first session with 0 points in terms of difficulty.}
    \begin{table}[h]
    \centering
    \footnotesize
    \caption{Difficulty points for the orientation experiment.}
    \begin{tabular}{llc}
    \toprule
    \textbf{Experiment}  & \textbf{Setting} & \textbf{Difficulty}\\
    \hline
    Orientation  & No Traffic + No Agent + No median + Day + Clear + Clear & 0\\
    \bottomrule
    \end{tabular}
    \end{table}

    \item {To represent the congested and free-flow conditions, the arrival rate and speed variables were combined into two categories. Congested (high arrival + low speed) and Free flow (low arrival + high speed). This further reduced the unique scenarios to 576.}
    
    \item {We considered 8 different scenarios (Table \ref{tab:Scenarios}) for the first pilot experiments and asked the participant to repeat these 8 sessions to investigate any learning effects. These scenarios were selected based on the distribution of difficulty level and representation of the three tiers in terms of importance.}

    \begin{table}[h]
    \centering
    \footnotesize
    \caption{Pilot configurations with difficulty levels.}
    \label{tab:Scenarios}
    \begin{tabular}{c p{3.75cm} p{2cm} l c }
    \toprule
    \textbf{Experiment} & \textbf{Tier 1}                                                 & \textbf{Tier 2}                      & \textbf{Tier 3}              & \textbf{Difficulty} \\ \hline
    1                   & 0\% AV + Low arrival + No agent + No signal                     & High speed + No median               & Day + Clear + Clear          & 1                   \\ \hline
    2                   & 0\% AV + Low arrival + Conservative + No signal                 & High speed + Median                  & Night + Clear + Clear        & 2                   \\ \hline
    3                   & 0\% AV + High arrival + No agent + No signal                    & Low speed + No median                & Day + Storm + Snow           & 3                   \\ \hline
    4                   & 0\% AV + High arrival + Conservative + No signal                & Low speed + Median                   & Night + Storm + Snow         & 4                   \\ \hline
    5                   & 100\% AV + Low arrival + No agent + Signal                      & High speed + No median               & Day + Clear + Clear          & 1                   \\ \hline
    6                   & 100\% AV + Low arrival + Conservative + Signal                  & High speed + Median                  & Night + Clear + Clear        & 2                   \\ \hline
    7                   & 100\% AV + High arrival + No agent + Signal                     & Low speed + No median                & Day + Storm + Snow           & 3                   \\ \hline
    8                   & 100\% AV + High arrival + Conservative + Signal                 & Low speed + Median                   & Night + Storm + Snow         & 4                   \\ \hline
    9                   & 100\% AV + Low arrival + No agent + No signal                   & High speed + No median               & Day + Clear + Clear          & 1                   \\ \hline
    10                  & 100\% AV + Low arrival + Conservative + No signal               & High speed + Median                  & Night + Clear + Clear        & 2                   \\ \hline
    11                  & 100\% AV + High arrival + No agent + No signal                  & Low speed + No median                & Day + Storm + Snow           & 3                   \\ \hline
    12                  & 100\% AV + High arrival + Conservative + No signal              & Low speed + Median                   & Night + Storm + Snow         & 4                   \\ \bottomrule
    \end{tabular}
    \end{table}

    \item {In the first pilot, we assessed the performance of the participants qualitatively and quantitatively.}
    \item {Based on the quantitative analysis, we further reduced the considered scenarios from 576 to 350 by randomly selecting categories from the 3rd tier variables for the generated scenarios.}
    \item {Based on the qualitative analysis, we realized that participants were gamifying the experimental task in the second round. Therefore, we decided to collect the data for one round only, but increased the number of sessions to 12.}
    \item {The second pilot was conducted to verify the design before conducting the main experiment.}

    \end{itemize}
    
\section{Additional Details of SCR Labelling Process}
\label{applabellingprocess}
\setcounter{figure}{0}
\renewcommand{\thefigure}{C\arabic{figure}}
\setcounter{table}{0}
\renewcommand{\thetable}{C\arabic{table}}

The process of labelling each detected SCR is explained in more detail in this section. There were four inputs required to perform the labelling process; 1) recording of the first-person-view of the resulting VR simulation, 2) the audio of the participant via the VR headset, 3) the figure of the EDA data with the identified SCRs for each session, and 4) the associated table of the identified SCRs for that session with a column to label each of the detected SCRs.

\begin{figure}[!ht] 
    \centering
    \includegraphics[width={0.885\linewidth}]{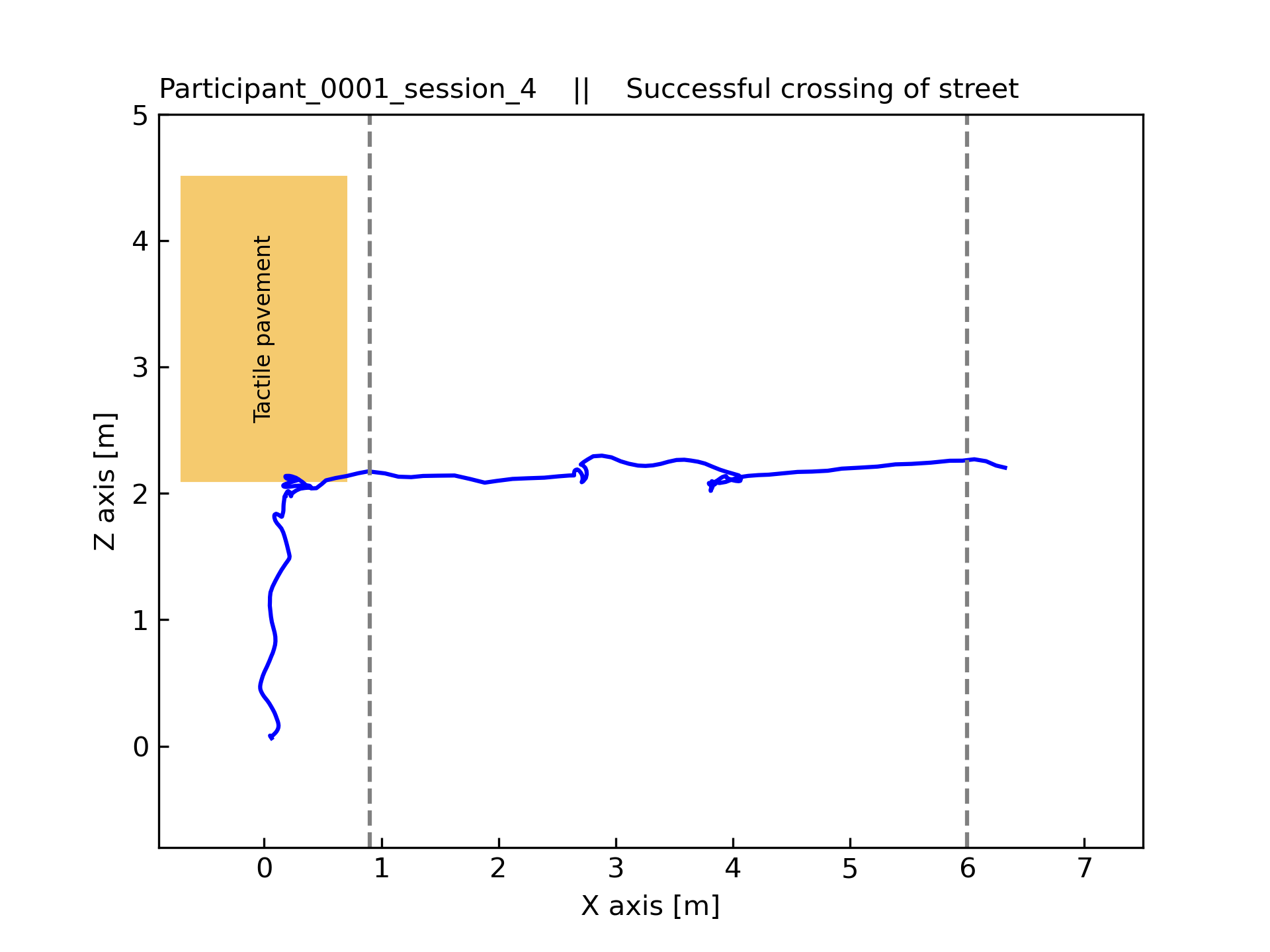}
    \caption{Trajectory of Participant 0001 in session 4}
    \label{fig:trajectory_participant_0001_session_4}
\end{figure}

\begin{figure}[!ht] 
    \centering
    \includegraphics[width={0.765\linewidth}]{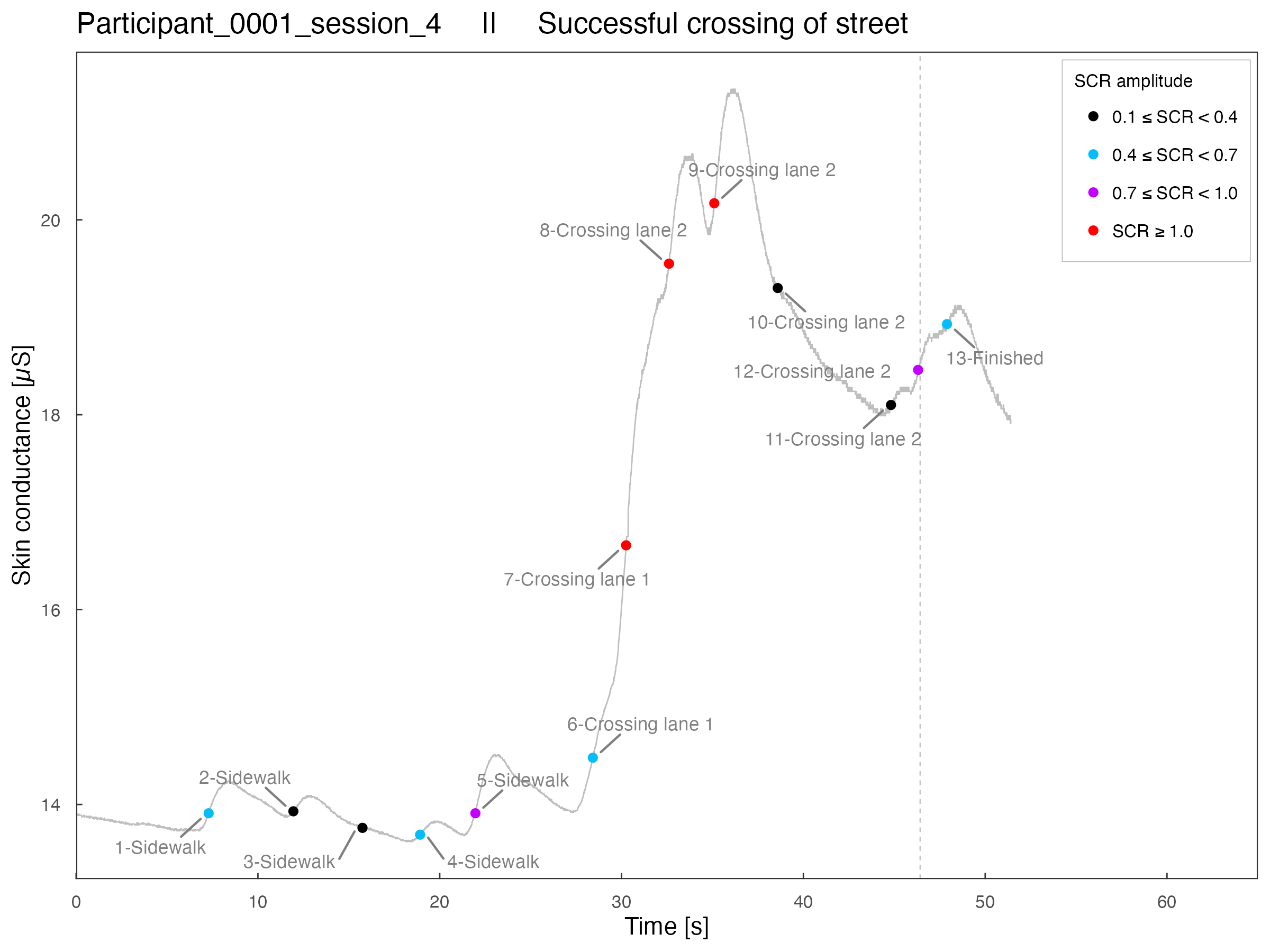}
    \caption{EDA signal and detected SCRs for Participant 0001 session 4}
    \label{fig:eda_participant_0001_session_4}
\end{figure}

Session 4 of participant 0001 has been selected to demonstrate the process. The video recording of this session can be found at \url{https://youtu.be/LdtyWLo07ts?si=REd7b_F0YZdOIr8A}. The trajectory of this participant in this session is shown in Figure \ref{fig:trajectory_participant_0001_session_4}. Figure \ref{fig:eda_participant_0001_session_4} shows the associated EDA signal and the identified SCRs for this session. Table \ref{tab:appendix} shows the details of the 13 detected SCRs in this session. The SCRs that were detected in the declining part of the signal and did not look like the typical arousal shape were labelled as ``Delete". Those SCRs that could not be associated to any of the pre-defined labels were labelled as ``Unknown".

\begin{landscape}
    
\begin{table}[!ht]
  \centering
  \vspace{5cm}
  \caption{Identified SCRs for Participant 0001 session 4}
    \begin{adjustbox}{max width=1.5\textwidth}
    \begin{tabular}{cllclcllllcl}
    \toprule
    \textbf{Participant\_id} & \textbf{Session\_id} & \textbf{Unix} & \textbf{Elapsed\_time} & \textbf{Position} & \textbf{SCR\_amplitude} & \textbf{SCR\_onset\_unix} & \textbf{SCR} & \textbf{Position\_f} & \textbf{Amp\_class} & \textbf{Detected\_SCR\_no} & \textbf{Annotation} \\
    \midrule
    0001  & Participant\_0001\_session\_4 & 1677282040.33 & 7.29  & Sidewalk & 0.50  & 1677282040.34 & 13.91 & Sidewalk & 0.4 $\leq$ SCR \textless 0.7 & 1     & Avatar's action \\
    0001  & Participant\_0001\_session\_4 & 1677282044.99 & 11.95 & Sidewalk & 0.31  & 1677282045.02 & 13.93 & Sidewalk & 0.1 $\leq$ SCR \textless 0.4 & 2     & Unknown \\
    0001  & Participant\_0001\_session\_4 & 1677282048.79 & 15.75 & Sidewalk & 0.19  & 1677282048.82 & 13.76 & Sidewalk & 0.1 $\leq$ SCR \textless 0.4 & 3     & Delete \\
    0001  & Participant\_0001\_session\_4 & 1677282051.96 & 18.92 & Sidewalk & 0.49  & 1677282052.01 & 13.69 & Sidewalk & 0.4 $\leq$ SCR \textless 0.7 & 4     & Unknown \\
    0001  & Participant\_0001\_session\_4 & 1677282054.99 & 21.96 & Sidewalk & 1.00  & 1677282055.00 & 13.91 & Sidewalk & 0.7 $\leq$ SCR \textless 1.0 & 5     & Avatar's action \\
    0001  & Participant\_0001\_session\_4 & 1677282061.46 & 28.42 & Crossing Lane 1 & 0.43  & 1677282061.49 & 14.48 & Crossing lane 1 & 0.4 $\leq$ SCR \textless 0.7 & 6     & Crossing \\
    0001  & Participant\_0001\_session\_4 & 1677282063.29 & 30.25 & Crossing Lane 1 & 3.42  & 1677282063.28 & 16.66 & Crossing lane 1 & SCR $\ge$ 1.0 & 7     & Checking the far-side traffic \\
    0001  & Participant\_0001\_session\_4 & 1677282065.65 & 32.61 & Crossing Lane 2 & 3.34  & 1677282065.68 & 19.55 & Crossing lane 2 & SCR $\ge$ 1.0 & 8     & Fear of accident \\
    0001  & Participant\_0001\_session\_4 & 1677282068.14 & 35.10 & Crossing Lane 2 & 2.71  & 1677282068.17 & 20.17 & Crossing lane 2 & SCR $\ge$ 1.0 & 9     & Fear of accident \\
    0001  & Participant\_0001\_session\_4 & 1677282071.63 & 38.59 & Crossing Lane 2 & 0.21  & 1677282071.67 & 19.30 & Crossing lane 2 & 0.1 $\leq$ SCR \textless 0.4 & 10    & Delete \\
    0001  & Participant\_0001\_session\_4 & 1677282077.86 & 44.82 & Crossing Lane 2 & 0.21  & 1677282077.86 & 18.10 & Crossing lane 2 & 0.1 $\leq$ SCR \textless 0.4 & 11    & Crossing \\
    0001  & Participant\_0001\_session\_4 & 1677282079.35 & 46.31 & Crossing Lane 2 & 0.76  & 1677282079.36 & 18.46 & Crossing lane 2 & 0.7 $\leq$ SCR \textless 1.0 & 12    & Crossing \\
    0001  & Participant\_0001\_session\_4 & 1677282080.94 & 47.90 & Finished & 0.68  & 1677282080.95 & 18.93 & Finished & 0.4 $\leq$ SCR \textless 0.7 & 13    & Crossing \\
    \bottomrule
    \end{tabular}%
    \end{adjustbox}
  \label{tab:appendix}%
\end{table}%
\end{landscape}

\newpage
\bibliographystyle{abbrvnat}
\singlespacing
\bibliography{references.bib}
\end{document}